\documentclass[useAMS,usenatbib]{mn2e}
\usepackage{apjfonts}
\usepackage{graphicx}
\usepackage{epstopdf}
\usepackage{amssymb}
\newcommand{\apj} {ApJ}
\newcommand{\aj} {AJ}
\newcommand{\apjl} {ApJL}
\newcommand{\aap} {AAp}
\newcommand{\apjs} {ApJS}
\newcommand{\aaps} {AApS}
\newcommand{\mnras}{MNRAS}
\newcommand{\pasp} {PASP}
\newcommand{\nat} {Nature}

\newcommand{\tab} {Table~}
\newcommand{\fig} {Figure~}
\newcommand{\sect} {Section~}

\title[Galaxy stellar mass functions and morphologies]{
Galaxy stellar mass functions of different morphological types in clusters,
and their evolution between z=0.8 and z=0}

\author[Vulcani et al. ]{\parbox[t]{\textwidth}{Benedetta Vulcani$^{1,2}$\thanks{E-mail:
benedetta.vulcani@oapd.inaf.it; }
\thanks{visiting The
Observatories of Carnegie Institution of Washington, Pasadena, CA, 
USA}\thanks{This file has been amended to highlight
the proper use of \LaTeXe\ code with the class file.  These changes
are for illustrative purposes and do not reflect the original paper by
B. Vulcani},
Bianca M. Poggianti$^{2}$, Alfonso Arag\'on-Salamanca$^{3}$, Giovanni Fasano$^{2}$, 
Gregory Rudnick$^{4}$, Tiziano Valentinuzzi$^{1}$, Alan Dressler$^{5}$, Daniela Bettoni$^{2}$, Antonio Cava$^{6,7}$, Mauro D'Onofrio$^{1}$,  
Jacopo Fritz$^{8}$, Alessia Moretti$^{2}$, Alessandro Omizzolo$^{2,9}$ and Jes\'us Varela$^{2}$}\\ 
\\
$^{1}$Astronomical Department, Padova University, Italy,\\ $^{2}$INAF-Astronomical Observatory of Padova, Italy,\\ $^{3}$School of Physics and Astronomy, University of 
Nottingham, Nottingham NG7 2RD, UK,\\ $^{4}$Department of Physics and Astronomy, University of Kansas, Lawrence, KS, USA,\\ $^{5}$The
Observatories of Carnegie Institution of Washington, Pasadena, CA, USA, \\$^{6}$Instituto de Astrofisica de Canarias, Spain,\\$^{7}$Departamento de Astrofisica, Universidad de La Laguna, Spain, \\$^{8}$Sterrenkundig Observatorium, Vakgroep Fysica en Sterrenkunde, Universiteit Gent, Belgium, \\$^{9}$Specola Vaticana, Vatican City, Hole See}

\begin{document}

\date{Accepted .... Received ..; in original form ...}

\pagerange{\pageref{firstpage}--\pageref{lastpage}} \pubyear{2010}

\maketitle

\label{firstpage}

\begin{abstract}
We present the galaxy stellar mass function and its evolution 
in clusters from $z\sim0.8$ to the
current epoch, based on 
the WIde-field Nearby Galaxy-cluster Survey (WINGS) ($0.04 \leq z \leq
0.07$), and the ESO Distant Cluster Survey
(EDisCS) ($0.4 \leq z \leq 0.8$).  We investigate the total mass
function and find it evolves noticeably with redshift.
The shape at $M_{\ast}> 10^{11} M_{\odot}$ does not evolve, 
but below $M_{\ast}\sim 10^{10.8} M_{\odot}$  the
mass function at high redshift is flat, while
in the Local Universe it flattens out at lower masses. 
The population of
$M_{\ast} = 10^{10.2} - 10^{10.8} M_{\odot}$ galaxies must have grown 
significantly between $z=0.8$ and $z=0$.
We analyze the mass functions of different morphological
types (ellipticals, S0s and late-types), and
find that also each of them 
evolves with redshift. All types have proportionally more massive galaxies
at high- than at low-z, and
the strongest evolution occurs among S0 galaxies.   
Examining the morphology-mass relation (the way the proportion of 
galaxies of different 
morphological types changes with galaxy mass), we find it strongly depends on
redshift.
At both redshifts, $\sim 40$\%  of the stellar mass is in elliptical galaxies.
Another $\sim 43$\% of the mass is in S0 galaxies
in local clusters,  while it is in spirals in distant clusters. 
To explain the observed trends, we
discuss the importance of those mechanisms that could shape 
the mass function.
We conclude that mass growth due to star formation plays a crucial
role in driving the evolution. It has to be accompanied by infall of
galaxies onto clusters, and the mass distribution of 
infalling galaxies might be different from that of cluster
galaxies. However, comparing with high-z field samples, we do not 
find conclusive evidence for such an environmental mass segregation. 
Our results suggest that star formation and infall change directly the mass
function of late-type galaxies in clusters and, indirectly, that of early-type
galaxies through subsequent morphological transformations.

\end{abstract}


\section{Introduction}
The distribution of galaxy stellar masses at the present day and in
the past is of fundamental importance for studying the assembly of
galaxies over cosmic time.

It is well known that galaxy properties depend both on galaxy
mass and on the environment the galaxies are part of (e.g. see
\citealt{baldry06, vanderwel08, bolzonella09, tasca09}).  Several
studies (e. g. \citealt{bundy05, vergani08}) have shown that mass
plays a crucial role in determining galaxy properties and in driving
their evolution. Color, specific star formation rate and internal
structure are strongly correlated with galaxy stellar mass
\citep{kauffmann03}. On the other hand, the stellar mass function of
galaxies depends on local galaxy density too (e.g. \citealt{baldry06,
bolzonella09}), showing that the mass distribution is itself related
to environment.  In addition, the efficiency of environmental
mechanisms can depend on the mass of the galaxy on which they act, and
this could cause environmental effects to be interpreted as
mass-effects \citep{tasca09}.

Clearly, galaxy mass and environment are strictly linked and it is
important to analyze them distinctly, trying to separate their
relative roles. To disentangle mass and environment it is useful to
study the dependence of some properties on mass fixing the
environment, or analyze the importance of the environment fixing the
mass.  In the nature vs nurture scenario, mass represents the primary
``intrinsic property'' closely related to primordial conditions, while
the environment includes all the possible external processes that can
influence galaxy evolution.

Stellar mass assembly is usually characterized by the galaxy stellar
mass function, that describes the distribution of galaxy masses, at
different epochs and different environments.  Several studies have
been carried out focusing on the importance of the galaxy mass
functions in the general field (e.g. \citealt{fontana06, drory05,
gwyn05, bundy06, pozzetti07}).  

For galaxies with $M \geq 10^{11}M_{\odot}$, several studies
(eg. \citealt{fontana04, bundy06, borch06}) have found that, overall,
 the evolution 
of the total mass function from $z=1$ to $z=0$
is relatively modest, which implies 
that the evolution of objects with mass
close to the local characteristic mass  is essentially complete
by z$\sim$1. However, \cite{fontana06} found that less massive 
galaxies evolve 
more than massive ones. Moreover, 
\cite{pozzetti09} showed a very fast rise of the total mass
function from z$\sim$1 for $M \leq 10^{11}M_{\odot}$ and 
they found this trend is due to 
 star formation driving the mass assembly at intermediate and
low masses. Instead, at high masses this process
does not play an important role and, as a consequence, \cite{pozzetti09}
found again that at high masses there is an almost negligible
evolution from $z\sim 1$ to $z\sim 0$.

It is interesting to investigate how galaxies of different types give
a different contribution to the mass function, shaping, for example,
either the massive tail or the low mass tail of the total mass
distribution.  There are several ways to subdivide galaxies into at
least two populations, usually either
according to their star formation histories
(being passive or star-forming, for example using a rest-frame color,
hence separating blue and red galaxies, their SEDs, their
spectroscopic features) or their structure (structural parameters or
morphologies).

In the Local Universe, a sort of bimodality in galaxy properties has
been observed: more massive galaxies tend to be passive, have high
surface mass densities, redder colors and have spectra characterized
by a pronounced break at $\lambda \sim$4000 \AA{}, while lower-mass
galaxies have low surface mass densities, tend to be still star
forming, show bluer colors and have spectra characterized by emission
lines \citep{kauffmann03, brinchmann04}. \cite{baldry04},
\cite{baldry06} and \cite{baldry08}, subdividing galaxies depending on
their colors, found a bimodal shape in the local mass function in the
field with an upturn related to the two different populations:
early-type galaxies dominate the high masses, while late-types
dominate at low/intermediate masses \citep{pozzetti09}.

Many studies have been performed to investigate the behavior of 
this bimodality at higher redshifts: for example, \cite{borch06}, 
who subdivide galaxies depending on 
their color,  \cite{fontana04}, \cite{bundy06} and \cite{vergani08},
who use spectral features to subdivide
galaxies with different properties. 
Although using different techniques, all of them emphasize
that 
the mass functions separately for early-type and late-type galaxies
depend on redshift. 
Early-type galaxies always dominate the massive end, while late-type galaxies
mostly contribute to the intermediate/low-mass part of the mass 
function at all redshifts. From $z\sim 1$ to $z\sim 0$,
early-type galaxies show a rapid evolution in their mass function,
while late-type galaxies show a relatively little change. 
This has been interpreted as the result of mass growth due to star formation
in late-type galaxies, accompanyed by a subsequent cessation of star formation
in some of the blue galaxies, that turn red and contribute to the build-up
of the red galaxy population \citep{bell07}.

Few studies have attempted to investigate the mass functions of
different galaxy types as a function of environment (see e.g. 
\citealt{balogh01, bundy06}).
\cite{bolzonella09}, classifying
galaxies following a spectral energy distribution classification, 
found that the shape of the mass function of
different galaxy types  is the same in
environments with different local density, so they argued for the
existence of a quasi-universal mass function for each type
regardless of environment, and also argued for a quasi-universal
evolution of the mass function of each type.  

The studies mentioned above all focus on the mass functions of
galaxies with different stellar histories in the general field, but
a clear picture on the stellar mass assembly and how
it depends on the mass (mass-assembly downsizing) and galaxy types has
not yet emerged. 

Fewer studies have focused on the
 mass functions as a function of galaxy morphology.  All of them consider
galaxies in the general field.
\cite{bundy05,
 pannella06, franceschini06} investigated mass functions of different
 morphological types, using ACS/HST images to classify galaxies. They
 found that, at $z\sim$1, morphologically early- and late-types 
 are present in similar numbers at all masses. By $z\sim
 0.3$, ellipticals dominate the high mass population, suggesting that
 merging or some other transformation process is active.  At all
 redshifts, they found that late-type galaxies dominate at lower
 masses, while early-type galaxies become prominent at higher
 masses.

Using both a spectral and a morphological classification, \cite{ilbert10}
found that most of massive quiescent galaxies have an elliptical
morphology at $z<$0.8. They also show similar mass functions.
Therefore, a dominant mechanism has 
to link the shutdown of star formation 
and the acquisition of an elliptical morphology in massive galaxies.

As the evolution of the mass functions of passive versus star-forming
galaxies must be influenced by the quenching of star formation in
previously star-forming galaxies (Bell et al. 2007), the evolution in
the mass functions of different morphological types must be influenced
by the transformation between $z=1$ and $z=0$ of some late-type
galaxies into early-type galaxies, of which evidence has accumulated
both in the field and, especially, in clusters.

In the field, \cite{bundy05} found evidence for the transformation of
late- into early-type galaxies as a function of time. Based on the
observed mass functions, this transformation process appears to be
more important at lower masses ($M_{\ast} \leq 10^{11} M_{\odot}$ )
because the most massive early-type galaxies are already in place at
z$\sim$1. Similarly, \cite{oesch09} showed that the evolution of the
morphological distribution in galaxies since $z\sim 1$ depends strongly
on stellar mass: ellipticals dominate the entire population since z=1, 
and the Hubble
sequence has quantitatively settled down by this epoch.  Most of the
morphological evolution from z=1 to z=0.2 takes place at masses
$M_{\ast} \leq 10^{11} M_{\odot}$, where the fraction of spirals
substantially drops and the contribution of early-types increases.
 
\cite{capak07} found that in the field galaxies are transformed from
late- to early-type galaxies more rapidly in dense than in sparse
regions.  Comparing their field data to cluster data, they found that
in clusters transformations occur even more rapidly.

The morphological evolution is much more striking in clusters, where
indeed was discovered: the fraction of late-type galaxies is much
higher, and the fraction of S0 galaxies proportionally lower, in
distant than in nearby clusters, suggesting an evolution from late to
early-type galaxies for a significant fraction of today's early-type
galaxies in clusters (Dressler et al. 1997, Fasano et al. 2000,
Postman et al. 2005, Smith et al. 2005, Desai et al. 2007).
Interestingly, the same studies find that the fraction of ellipticals
in clusters does not evolve with redshift.

It isn't surprising that the impact of morphological transformations
depends on environment.
In clusters, galaxies are expected to be affected by a number of
physical processes that can alter the course of their evolution.
Galaxies in dense environments are subject to a variety of external
stresses, which are, in general, not conductive to the maintenance of
spiral structure: i.e. ram pressure \citep{gunn72, bekki09}, galaxy
harassment \citep{moore96}, cluster tidal forces \citep{byrd90},
interaction/merging \citep{icke85, bekki98}, strangulation
\citep{larson80, font08, mccarthy08} can act with different efficiency
depending on environment, so field galaxies infalling into larger
structures can be
transformed from gas-rich spirals to gas-poor lenticular galaxies.

Until now, a few studies of the mass function of galaxies in
clusters has been carried out, due to the few observations available.
 \cite{kodama03} have estimated the stellar mass function for two
clusters at $z\sim1$ transforming their K$_s$-band luminosity function
by correcting it for the star formation contribution as estimated from
the J-K$_s$ colours. This and other works (e.g. \citealt{aragon93, barger98, toft04, andreon06, 
depropris07}) have highlighted the fact that the
evolution of the characteristic luminosity $M_{\star}$ at red
wavelengths in clusters is consistent with passive evolution and with
massive cluster galaxies being fully assembled by $z \sim 1$. In
particular, \cite{toft04} pointed out an evolution in the faint
end slope of the K$_s$-band luminosity function of one cluster at
$z=1.2$, suggesting that clusters
at $z \sim 1$ contain relatively smaller fractions of low mass galaxies
than clusters in the local Universe.
%

In this context, the WIde-field
Nearby Galaxy-cluster Survey (WINGS) \citep{fasano06} and the ESO
Distant Cluster Survey (EDisCS) \citep{white05}, two surveys whose
main aims are to characterize clusters and their galaxies at low and
high redshift respectively, represent a significant increase in the
number of well studied clusters at low and high redshift respectively.

In this paper we focus on the evolution of the galaxy mass functions in clusters.
We first analyze the total mass function, and then
separately the mass functions of different morphological types 
and the evolution of the morphological fractions, 
with the aim of understanding the mechanisms that drive the evolution of the 
mass distributions. 

This paper is organized as follows. In \S 2 we present the data-sets
used (WINGS and EDisCS), describing the surveys and the clusters
selection at different redshifts; in \S 3 we derive stellar masses. In
\S 4 we present the mass and magnitude limited
galaxy samples, and other galaxy properties we need
for the analysis. 
We then
show the total mass function and its evolution with redshift (\S5.1), and
the 
mass functions of different galaxy morphological types (\S5.2) and their
evolution from $z \sim 0.8$ to the current epoch (\S5.3).  In \S 6 
we analyze the morphology-mass relation,
the evolution of the morphological fractions and of the
mass fractions in each morphological type.
In \S 7 we comment our findings, discussing the possible
mechanisms that are responsible for the observed evolution. Finally, 
in \S 8 we summarize our conclusions.
 
Throughout this paper, we assume $H_{0}=70 \, \rm km \, s^{-1} \, Mpc^{-1}$, 
$\Omega_{m}=0.30$, $\Omega_{\Lambda} =0.70$.  The adopted initial 
mass function (IMF) is a \citet{kr01} in the mass range 0.1-100 $M_{\odot}$.

\section{The data samples}
To perform the study of the mass distribution of galaxies 
of different morphological types and its evolution
from $z\sim 0.8$ to the current epoch, 
we assemble two galaxy cluster samples in two redshift intervals: 
the sample at low-z is selected from the WIde-field Nearby Galaxy-cluster 
Survey (WINGS) \citep{fasano06}, while the sample at high-z is selected 
from the ESO Distant Cluster Survey (EDisCS) \citep{white05}.

\subsection{Low-z sample: WINGS}\label{samplew}
WINGS\footnote{http://web.oapd.inaf.it/wings}  \citep{fasano06} 
is a multiwavelength survey of clusters at \mbox{$0.04 < z < 0.07$}
whose main goal is to characterize the photometric and
spectroscopic properties of galaxies in nearby clusters and
to describe the changes of these properties depending on 
galaxy mass and environment.

Clusters were selected in the X-ray from the ROSAT 
Brightest Cluster Sample and its extension \citep{ebeling98, ebeling00}
 and the X-ray Brightest Abell-type Cluster sample \citep{ebeling96}. 
 WINGS clusters cover a wide range of velocity dispersion 
 $\sigma_{clus}$ (typically $500-1100 \, km \, s^{-1}$) and a 
 wide range of  X-ray luminosity $L_{X}$ (typically $0.2-5 \times10^{44} erg \, s^{-1}$).

The survey is mainly based on optical B, V imaging of 78 galaxy
clusters \citep{varela09}, that has been complemented by a
spectroscopic survey of a subsample of 48 clusters, obtained with the
spectrographs WYFFOS@WHT and 2dF@AAT \citep{cava09}, by a
near-infrared (J, K) survey of a subsample of 28 clusters obtained
with WFCAM@UKIRT \citep{valentinuzzi09}, and by U broad-band and
$H_{\alpha}$ narrow-band imaging of a subset of WINGS clusters,
obtained with wide-field cameras at different telescopes (INT, LBT,
Bok) (\citealt{omizzolo10}).

The spectroscopic target selection was based on the WINGS 
B, V photometry. The aim of the target
selection strategy was to maximize the chances of observing
galaxies at the cluster redshift without biasing the cluster sample.
We selected galaxies with a total $V \leq 20$ magnitude and a V magnitude
within the fiber aperture of V $<$ 21.5 and with a color within a 
5 kpc aperture of $(B-V)_{5 kpc} \leq1.4$, to reject background
galaxies, much redder than the cluster red sequence. 
The exact cut in color was varied slightly from cluster
to cluster in order to account for the redshift variation and
to optimize the observational setup. 
These very loose selection limits were applied so as to avoid any bias
in the colors of selected galaxies. 

In this paper, we only consider spectroscopically confirmed members of
21 of the 48 clusters with spectroscopy.  This is the subset of
clusters that have a spectroscopic completeness (the ratio of the
number of spectra yielding a redshift to the total number of galaxies
in the photometric catalog) larger than 50\%.  The completeness is
determined using V-band magnitudes and 
turns out to be
rather flat with magnitude for most clusters, for the reasons discussed
in \cite{cava09}. Moreover, completeness
is essentially independent from the distance to the centre of the
cluster for most clusters.

Our imaging covers a
 $34^{\prime} \times 34^{\prime}$ field, that corresponds to about 0.6
 $R_{200}$, for all the 21 clusters considered, except for
{\it A1644} and {\it A3266} where the $\sim 0.5 R_{200}$ is covered.
$R_{200}$  is defined as the radius delimiting a 
sphere with interior mean density 200 times the critical 
density of the Universe at that redshift,
and is commonly used as an approximation for the cluster 
virial radius. The $R_{200}$  values for our structures are computed 
from the velocity dispersions by \cite{cava09}.

Galaxies are considered members of a cluster if their spectroscopic
redshift lies within $\pm3 \sigma$ from the cluster mean redshift,
where $\sigma$ is the cluster velocity dispersion \citep{cava09}.

The clusters used in this analysis are listed in \tab\ref{tab:wi.cl}.  

\begin{table}
\centering
\begin{tabular}{cccc}
\hline
cluster name & z & $\sigma$  		& DM\\
	   &	& ($km \, s^{-1}$)		&  {\it (mag)}\\
\hline
A1069 & 0.0653 & 690$\pm$ 68 & 37.34\\
A119 & 0.0444 & 862$\pm$ 52& 36.47\\
A151 & 0.0532 & 760$\pm$55 &36.87 \\
A1631a & 0.0461 & 640$\pm$33 &36.55 \\
A1644 &0.0467 & 1080$\pm$ 54& 36.58\\
A2382 & 0.0641 & 888$\pm$ 54&37.30 \\
A2399 & 0.0578 & 712$\pm$ 41& 37.06\\
A2415& 0.0575 & 696$\pm$ 51& 37.05\\
A3128& 0.06 & 883$\pm$ 41& 37.15\\
A3158& 0.0593 & 1086$\pm$ 48&37.12 \\
A3266& 0.0593 & 1368$\pm$ 60&36.12 \\
A3376 &0.0461 & 779$\pm$ 49&36.55 \\
A3395 & 0.05 & 790$\pm$ 42&36.73 \\
A3490 & 0.0688 & 694$\pm$ 52&37.46 \\
A3556 & 0.0479 & 558$\pm$ 37&36.64 \\
A3560 & 0.0489  & 710$\pm$ 41&36.68 \\
A3809 & 0.0627 & 563$\pm$ 40&37.25 \\
A500 &0.0678  & 658$\pm$48 & 37.42 \\
A754 & 0.0547 & 1000$\pm$ 48& 36.94\\
A957x & 0.0451 &710$\pm$ 53& 36.50\\
A970 & 0.0591 & 764$\pm$ 47&37.11\\
\hline
\end{tabular}
\caption{List of WINGS clusters analyzed in this paper and their redshift $z$, their velocity dispersion 
$\sigma$ and their distance modulus $DM$. \label{tab:wi.cl}  }
\end{table}

\subsection{High-z sample: EDisCS} \label{dataediscs}
EDisCS is a multiwavelength photometric and spectroscopic survey 
of galaxies in 20 fields containing galaxy clusters at 
$0.4< z <1$ \citep{white05}. Its main goal is to characterize
both the clusters themselves and the galaxies within them.

Clusters were drawn from the Las Campanas Distant
Cluster Survey (LCDCS) catalog \citep{gonzalez01}. They were selected
as surface brightness peaks in smoothed images taken with a very wide
optical filter ($\sim 4500-7500$ \AA{}). The redshifts of the LCDCS cluster 
candidates were initially estimated from the apparent magnitude
 of the brightest cluster galaxy (BCG).  The 20 EDisCS fields 
 were chosen among the 30 highest surface brightness 
candidates, after confirmation of the presence of an 
apparent cluster and of a possible red sequence with VLT 20 
min exposures in two filters \citep{white05}. Then, all candidates were subdivided in 
 two groups, the first one at intermediate redshift ($z\sim 0.5$) and
 the second one at high redshift ($z\sim 0.8$).

For all 20 fields, EDisCS has obtained deep optical multiband
photometry with FORS2/VLT \citep{white05} and near-IR photometry with
SOFI/NTT \citep{aragon09}.  Photometric redshifts were measured using
both optical and infrared imaging \citep{pello09}.  Their
determinations, their performance, and their use to isolate cluster
members, are described in detail in \cite{pello09} and
\cite{rudnick09}, and are briefly summarized in the following.
Photometric redshifts were computed for every object in the EDisCS
fields using two independent codes, a modified version of the publicly
available Hyperz code \citep{bolzonella00} and the code of
\cite{rudnick01} with the modifications presented in
\cite{rudnick03}. The accuracy of both methods is $\sigma (\delta z)
\sim 0.05-0.06$, where $\delta z = \frac{zspec-zphot} {1+zspec}.$
Membership was established using a modified version of the technique
first developed in \cite{brunner00}, in which the probability of a
galaxy to be at redshift $z$ ($P(z)$) is integrated in a slice around
the cluster redshift to give $P_{clust}$ for the two codes. The width
of the slice around which $P(z)$ is integrated should be of the order
of the uncertainty in redshift for the galaxies in question. In our
case a $\Delta z = \pm 0.1$ slice around the spectroscopic redshift of
the cluster $z_{clust}$ was used. A galaxy was rejected from the
membership list if $P_{clust}$ was smaller than a certain probability
$P_{thresh}$ for either code.  The $P_{thresh}$ value for each cluster
was calibrated from our spectroscopic redshifts and was chosen to
maximize the efficiency with which we can reject spectroscopic
non-members while retaining at least $\sim 90\%$ of the confirmed
cluster members, independent of their rest-frame (B-V) color or
observed (V-I) color. In practice we were able to choose thresholds
such that we satisfied this criterion while rejecting 45\%-70\% of
spectroscopically confirmed non-members. Applied to the entire
magnitude limited sample, our thresholds reject 75\%-93\% of all
galaxies with $I_{tot}$ < 24.9.  A posteriori, we verified that in
our sample of galaxies with spectroscopic redshift and above the mass
limit described below, $\sim$20\% of those galaxies that are
photo-z cluster members are spectroscopically interlopers and,
conversely, only $\sim$6\% of those galaxies that are spectroscopic
cluster members are rejected by the photo-z technique.

Deep spectroscopy with FORS2/VLT was obtained for 18 of the fields
\citep{halliday04, milvang08}. 
Spectroscopic targets were selected with the aim 
to produce an unbiased sample of cluster galaxies. 
Conservative rejection criteria based on photometric redshifts
(different from those adopted to assess the photo-z membership
described above) were used in the selection of spectroscopic targets
to reject a significant fraction of non-members, while retaining
an unbiased spectroscopic sample equivalent to a purely I-band selected one.
Typically, spectra of more than 100 galaxies per field were obtained,
with I$\leq$22 for the z$\sim$0.5 cluster candidates and
I$\leq$23 for the z$\sim$ 0.8 ones.

ACS/HST mosaic imaging in $F814W$ 
of 10 of the highest redshift clusters has also been acquired 
\citep{desai07}, covering with four ACS pointings a $6.5^{\prime} \times6.5^{\prime}$ 
field with an additional deep pointing in the centre. This field covers
the $R_{200}$ of all clusters, except for {\it cl 1232.5-1250}
\citep{poggianti06}. The $R_{200}$ values for our
structures are computed from the velocity dispersions by
\cite{poggianti08}. 

In this paper, to have a large sample of galaxies at high redshift (see below), 
we analyze the data of photo-z members of 9 EDisCS
clusters, for which {\it HST} images and, consequently, morphologies
were available.  Even though {\it HST}
data for {\it cl 1037.7-1243} are available, 
this cluster  has no galaxies above
the completeness limits we will adopt, hence in our analysis we do not consider it.
Table \ref{tab:ed_cl} presents the list of clusters
used.

\begin{table}
\centering
\begin{tabular}{|c|c|c|}
\hline
cluster name & z & $\sigma$  \\
	   &	& ($km \, s^{-1}$)\\
\hline
cl 1040.7-1155 & 0.70 &418$^{+55}_{-46}$\\
cl 1054.4-1146 & 0.70 &589$^{+78}_{-70}$\\ 
cl 1054.7-1245 &0.75 &504$^{+113}_{-65}$ \\
cl 1103.7-1245    &0.62 &336$^{+36}_{-40}$\\
cl 1138.2-1133    &0.48 &732$^{+72}_{-76}$\\
cl 1216.8-1201 & 0.79 &1018$^{+73}_{-77}$\\
cl 1227.9-1138&0.64 &574$^{+72}_{-75}$\\
cl 1232.5-1250 & 0.54&1080$^{+119}_{-89}$\\
cl 1354.2-1230    & 0.76 &648$^{+105}_{-110}$\\
\hline
\end{tabular}
\caption{List of EDisCS clusters analyzed in this paper, with cluster name, 
redshift $z$ and velocity dispersion $\sigma$ (from  \citealt{halliday04, milvang08}).
\label{tab:ed_cl}}
\end{table}

\section{Galaxy stellar masses estimates}\label{sec:mass}
In this paper we adopt the \cite{kr01} IMF in the mass range 0.1-100 $M_{\odot}$ to
compute the galaxy stellar masses.  
We determine them in the same way for both
data-sets.  
We follow \cite{bj01} who used a spectrophotometric
model finding a strong
correlation between stellar mass-to-light ($M/L$) 
ratio and opticals colors of the
integrated stellar populations for a wide range of star formation histories. 
They found that this correlation is quite robust to 
uncertainties in stellar population and galaxy evolution
models. 

We determine stellar masses using the relation 
 between  $M/L_{B}$  and rest-frame $(B-V)$ color
and the equation given in \cite{bj01}:
\begin{equation}
\log_{10}(M/L_{B})=a_{B}+b_{B}(B-V)
\end{equation}
For the Bruzual \& Charlot model with a \cite{salpeter55} IMF 
(0.1-125 $M_{\odot}$) and
solar metallicity, $a_B=-0.51$ and $b_B= 1.45$. 
Then, we scale our masses  to a \cite{kr01} IMF adding -0.19 dex 
to the logarithmic value of the masses.

The choice of the photometric bands is dictated by the fact that
we have only B and V for all the WINGS clusters used,
and also that the rest-frame B-band 
has been directly observed for all EDisCS clusters.
K-band luminosity would be preferable to determine masses, as the K-band is
less sensitive to recent star formation and dust. However, K-band is available
only for a subset of the WINGS clusters considered, and 
K-band wavelengths have not been directly observed by EDisCS, for which 
K-band magnitudes 
were extrapolated from the photo-z templates. 

\subsection{Low-z: WINGS}
\begin{figure}
\centering
\includegraphics[scale=0.43]{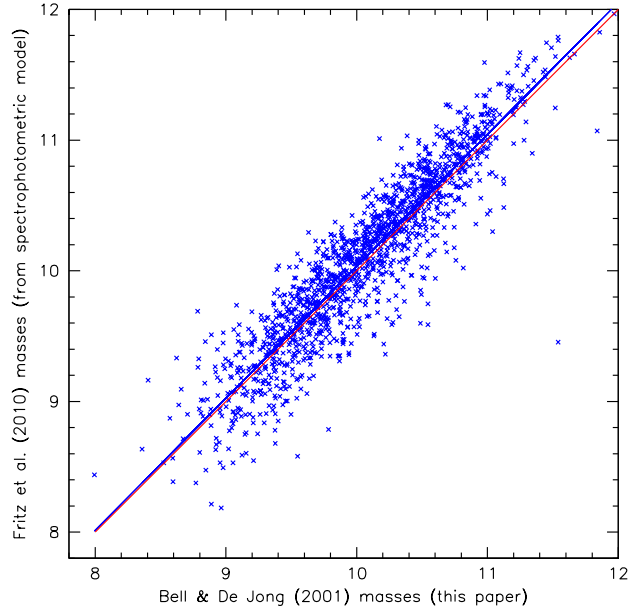}
\caption{Comparison between the determination of WINGS masses used in this 
paper and those derived from
the spectrophotometric model \citep{fritz07, fritz10}. All masses 
are converted to a \citet{kr01} IMF in the mass range 0.1-100 $M_{\odot}$.  Points 
are logarithmic values. The blue line is the best fit to the points, and the red line is the 1:1 
relation. \label{massew}}
\end{figure} 

For WINGS spectroscopically confirmed cluster galaxies, the total
luminosity $L_{B}$ has been derived from the total (Sextractor AUTO)
observed B magnitude \citep{varela09}, corrected for distance modulus
and foreground Galaxy extinction, and k-corrected using tabulated
values from \cite{poggianti97}.
The B-V color used to calculate masses was
derived from observed B and V aperture magnitudes
measured within a diameter of 10 kpc around each galaxy baricenter, 
corrected as the total magnitude.
 
Stellar masses for WINGS galaxies had been previously determined
by fitting the optical spectrum (in the range \mbox{$\sim 3600 - \sim 7000$} 
\AA{}) (\citealt{fritz10}), with the spectro-photometric 
model fully described in 
\cite{fritz07}. In this model, all the main 
spectro-photometric features are reproduced by 
summing the theoretical spectra of Simple Stellar Population 
(SSP) of 13 different ages (from $3\times 10^{6}$ to $\sim
14\times 10^{9}$ years). 
Dust extinction is allowed to vary as a function of SSP age, 
and the metallicity can vary among three values: Z=0.004, Z=0.02 and Z=0.05.
These mass estimates 
were then corrected for color gradients within each galaxy (i.e.
for the difference in color within the fibre and over a 10kpc diameter).
For a detailed description 
of the determination of masses see \cite{valentinuzzi09sd} and
\cite{fritz10}.

The comparison of masses used in this paper with masses determined 
by the \cite{fritz10} spectro-photometric model converted to the adopted IMF
(see Fig.~\ref{massew}) shows no offset
between the two estimates, and an rms of  $\sim 0.3$ dex that we adopt 
as the typical error on the masses.  
Above $\log M_{\ast}/M_{\odot}=9.8$, that is the
WINGS mass completeness limit (see \S \ref{w_m}) there is an offset of $\sim 0.06$ dex and the rms 
is  $\sim 0.3$ dex.
An exhaustive comparison among masses and errors determined
from different bands can be found in \cite{fritz10}.

For homogeneity with the EDisCS
mass estimates (see next section), throughout this paper we only use
\cite{bj01} masses.

\subsection{High-z: EDisCS}

\begin{figure}
\centering
\includegraphics[scale=0.43]
{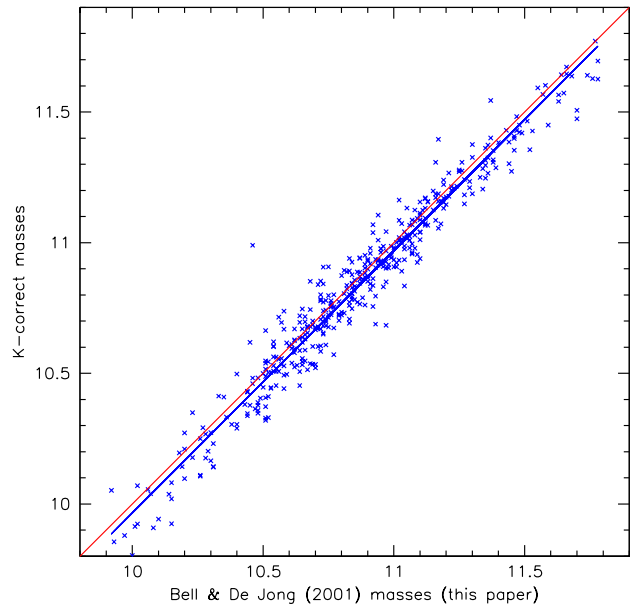}
\caption{Comparison between the determination of EDisCS masses used in this 
paper and those derived using the {\it k-correct} tool \citep{blanton07}.
Only galaxies for which spectroscopy is available are considered. All masses 
are converted to a \citet{kr01} IMF in the mass range 0.1-100 $M_{\odot}$.  Points 
are logarithmic values.
The blue line is the best fit to the points, and the red line is the 1:1 
relation. \label{massee}}
\end{figure}

For EDisCS galaxies, 
stellar masses for spectroscopic members were estimated by John Moustakas
(2009, private communication) 
using the {\it kcorrect} tool \citep{blanton07}{\footnote{http://cosmo.nyu.edu/mb144/kcorrect/}, that
models the available observed broad band photometry, fitting templates
obtained with spectrophotometric models.

For this paper, to be fully consistent with stellar masses of the sample
at low-z, we have estimated stellar masses following again  the 
method proposed by \cite{bj01}, using total 
absolute magnitudes derived from photo-z fitting fixing each
galaxy redshift to be equal to
the spectroscopic redshift of the cluster it belongs to
\citep{rudnick03, pello09},
converting our values to a \cite{kr01} IMF.

In \fig\ref{massee}, a comparison of the two methods is shown
 only for spectroscopic members of all EDisCS clusters (not just the 9
with HST used in this paper). There is a satisfactory agreement,
with a systematic offset of 0.03 dex and an rms of 0.08 dex.
If we consider data only above the EDisCS completeness limit 
($\log M_{\ast}/M_{\odot} \geq 10.4$, see \S \ref{e_m}), 
we find a mean offset of $\sim 0.09$ dex and an rms of 0.08 dex.

Finally, we note that a only slightly larger systematic offset
  and rms (0.13 and 0.13, respectively) are found when comparing the
  masses of EDisCS spectroscopic members derived with the \cite{bj01}
  method with those derived by fitting the EDisCS spectra using the
  non-parametric method of \cite{ocvirk06, ocvirk10}, the new models of
  \cite{vazdekis09} and the MILES stellar library of \cite{sb06}.

\section{Galaxy samples}

Both at high- and low-z, we analyze the mass functions and the
fractions of galaxies of different morphological types, as determined
by high-quality imaging.  For this purpose, galaxies are divided in
three groups: ellipticals, lenticulars (or S0s) and late-type galaxies
(spirals + irregulars). In some cases, we will also consider
ellipticals and S0s together, defined as early-type galaxies.

In the following, depending on 
our aim, we use two different methods of galaxy sample selection.

To study the mass distribution of different morphological types (\S5), 
the best is to use a mass limited sample.
In this case, the mass limit endorsed needs to be such to
ensure completeness, i.e. to include all galaxies more massive than the limit
regardless of their color or morphological type.
To determine this limit, we compute the mass of an object whose
observed magnitude is equal to the faint magnitude limit
of the survey, and whose color is the reddest color 
of a galaxy at the highest redshift considered. 

In the subsequent part of the paper (\S6), we compare the evolution of the
mass distributions with the evolution of the morphological fractions.
Since in the literature the latter are available only for a 
magnitude limited sample, we assemble also 
galaxy magnitude limited samples.
In this case, the minimum detected stellar
mass depends both on redshift and galaxy color.

\subsection{Low-z sample: WINGS}
For WINGS, we consider only spectroscopically confirmed members of our clusters.
Using all galaxies in the photometric sample would provide a larger dataset,
however for WINGS the 
photometric data cannot be used to assess cluster membership
due to the low redshift and restricted photometric coverage.

Since galaxies with available spectroscopy are only a fraction
($>50\%$ in the clusters used) of the total number of possible 
spectroscopic targets, we need to apply a statistical correction to
correct for incompleteness. This is obtained weighting each galaxy by
the inverse of the ratio of the number of spectra yielding a redshift
to the total number of galaxies in the photometric catalog, in bins of
1 mag \citep{cava09}.

 As in several other works (see e.g. \citealt{Linden10}), in each
cluster, we exclude the BCG, defined as the most luminous galaxy of
each cluster, that could alter the mass distribution of the whole
sample.  In fact, the properties of BCGs are in many respect peculiar
with respect to other galaxies, and they are the subject of many
separate studies dedicated only to this class of objects. For example,
\cite{fasa10} found that also in the WINGS dataset BCGs are
heterogeneus with respect to the global population of ellipticals for
their distribution of apparent (and intrinsic) flattenings.  

Only galaxies lying within 0.6$R_{200}$ are considered, because this is
the largest radius covered approximately in all clusters.

Morphological types are derived from V-band images using MORPHOT, an
automatic tool for galaxy morphology, purposely devised in the
framework of the WINGS project. 
MORPHOT was designed with the aim to reproduce as closely as possible
visual morphological classifications.

MORPHOT extends the classical CAS
(Concentration/Asymmetry/clumpinesS) parameter set \citep{conselice03}, by
using 20 image-based morphological diagnostics. Fourteen of them have
never been used, while the remaining six [besides the CAS parameters,
the Sersic index, the Gini and M20 coefficients \citep{lotz04}] are
actually already present in the literature, although in slightly
different forms.
Defining the newly introduced diagnostics and explaining their meaning
is beyond the scope of the present paper. We refer the reader to
\citet[][Appendix~A therein]{fasa10} for an outlining of the logical
sequence and the basic procedures of MORPHOT, while an exhaustive
description of the tool will be given in a forthcoming paper (Fasano
et al., in preparation). Here we just mention that, among the 14 newly
devised diagnostics, the most effective one in order to disentangle ellipticals
from S0 galaxies turned out to be an Azimuthal coefficient, measuring the
correlation between azimuth and pixel flux relative to the average
flux value of the elliptical isophote passing through the pixel
itself. Figure~\ref{morph1} illustrates the distributions of the
Azimuthal coefficient for the samples of ellipticals and S0 galaxies used to 
calibrate MORPHOT.  

More importantly for our purposes, 
the quantitative discrepancy between automatic (MORPHOT)
and visual classifications turns out to be similar to the typical
discrepancy among visual classifications given by experienced, independent
human classifiers (r.m.s.$\sim$1.5-2.5 T types). 
This was extensively tested on a calibration sample of 931 WINGS galaxies
with both visual and MORPHOT classification, where the visual classification
was carried out independently by GF and AD. This is illustrated in
Figure~\ref{morph2}, where the automatic classification is also shown
to be bias-free in the overall range of morphological types.

\begin{figure}
\includegraphics[scale=0.35]{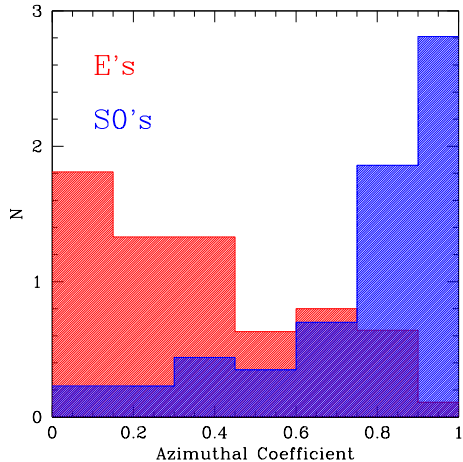}
\caption{
Normalized distributions of the MORPHOT Azimuthal Coefficient for the
visually classified ellipticals (366 objects, red histogram) and S0
galaxies (267 objects, blue histogram) of the MORPHOT calibration
sample.
\label{morph1}}
\end{figure}

\begin{figure}
\includegraphics[scale=0.35]{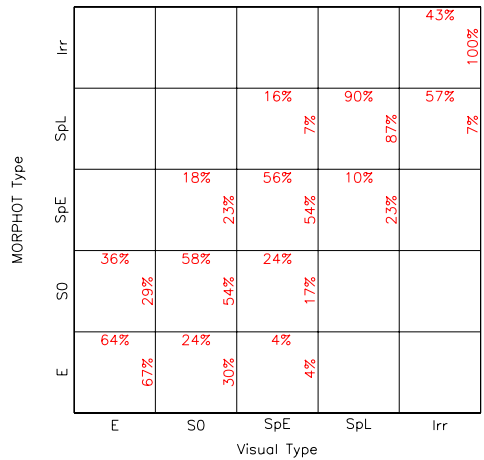}
\caption{Comparison between visual and MORPHOT broad morphological types
for the 931 galaxies of the MORPHOT calibration sample. In the 2D bins
of the plot the percentages of visual/MORPHOT broad types in different
MORPHOT/visual types are reported in the top/right sides of the bins.   
\label{morph2}}
\end{figure}

For now, we can apply MORPHOT just to the WINGS imaging,
because the tool is calibrated on the WINGS imaging characteristics,
and we defer to a later time a more generally usable version of the
tool. To verify directly that the two methods adopted at different
redshifts (see \sect\ref{sample_ed})
are consistent, we can apply
the same ``method'' (visual classification and persons)
that was used at high-z on the low-z images.

To this aim, 3 of the classifiers that in 2007 visually classified all
the EDisCS galaxies (BMP, AAS, VD) now performed a visual
classification of WINGS galaxies. This was done on the subset of
WINGS galaxies that was used to calibrate MORPHOT on the visual WINGS
morphologies, including only galaxies that enter
the sample we analyze in this paper (173 galaxies).

The results (see Fig.\ref{mor}) show agreement between the three broad
morphological classes assigned by the EDisCS classifiers with the
WINGS visual classification in $\sim$83\% of the
cases, and with MORPHOT in $\sim$75\% of the cases.


\begin{figure*}
\begin{minipage}[c]{175pt}
\centering
\includegraphics[scale=0.35,clip = false, trim = 0pt 150pt 0pt 0pt]
{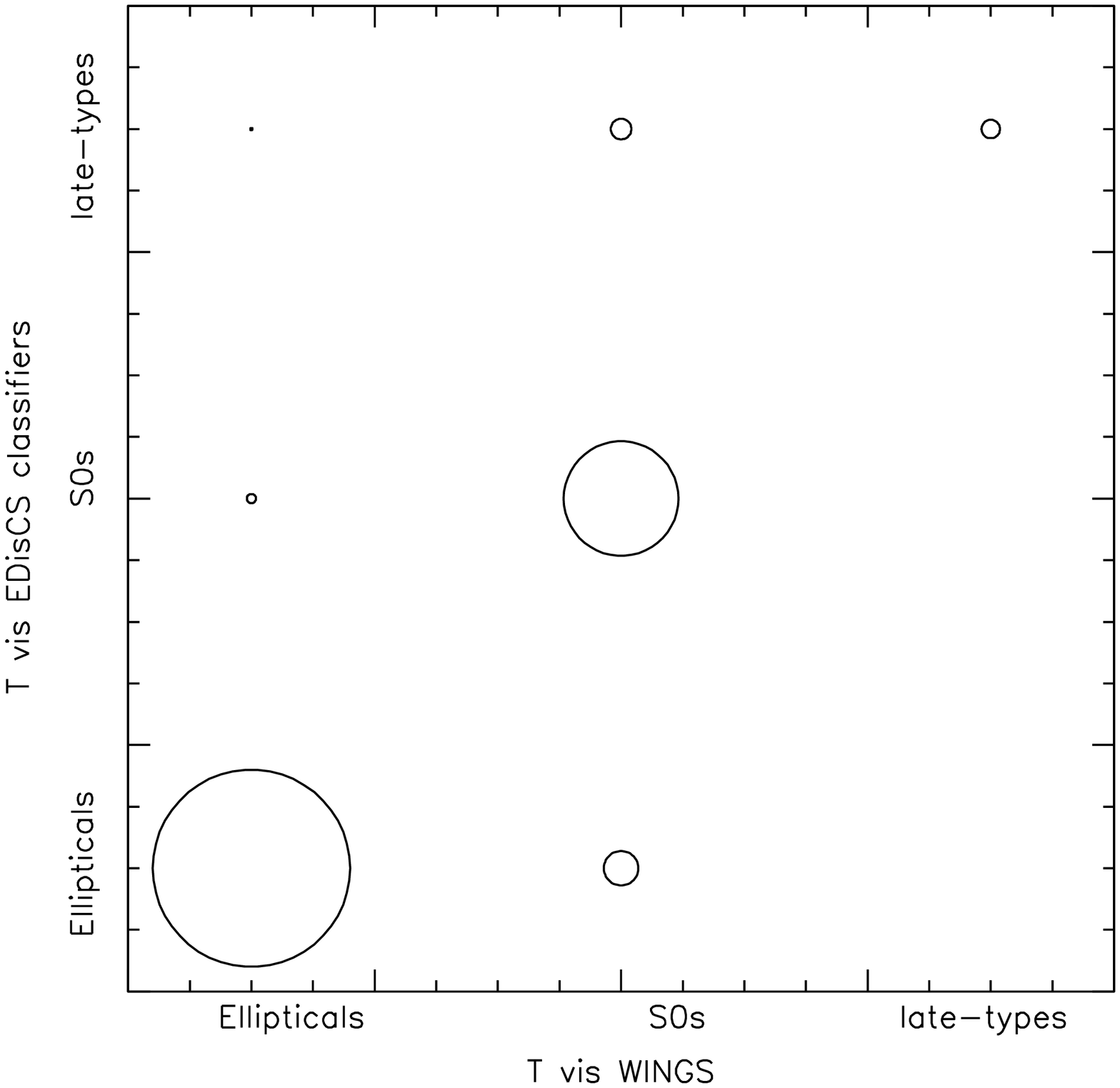}
\vspace*{2.2cm}
\end{minipage}
\hspace{2.5cm}
\begin{minipage}[c]{175pt}
\centering
\includegraphics[scale=0.35,clip = false, trim = 0pt 150pt 0pt 0pt]
{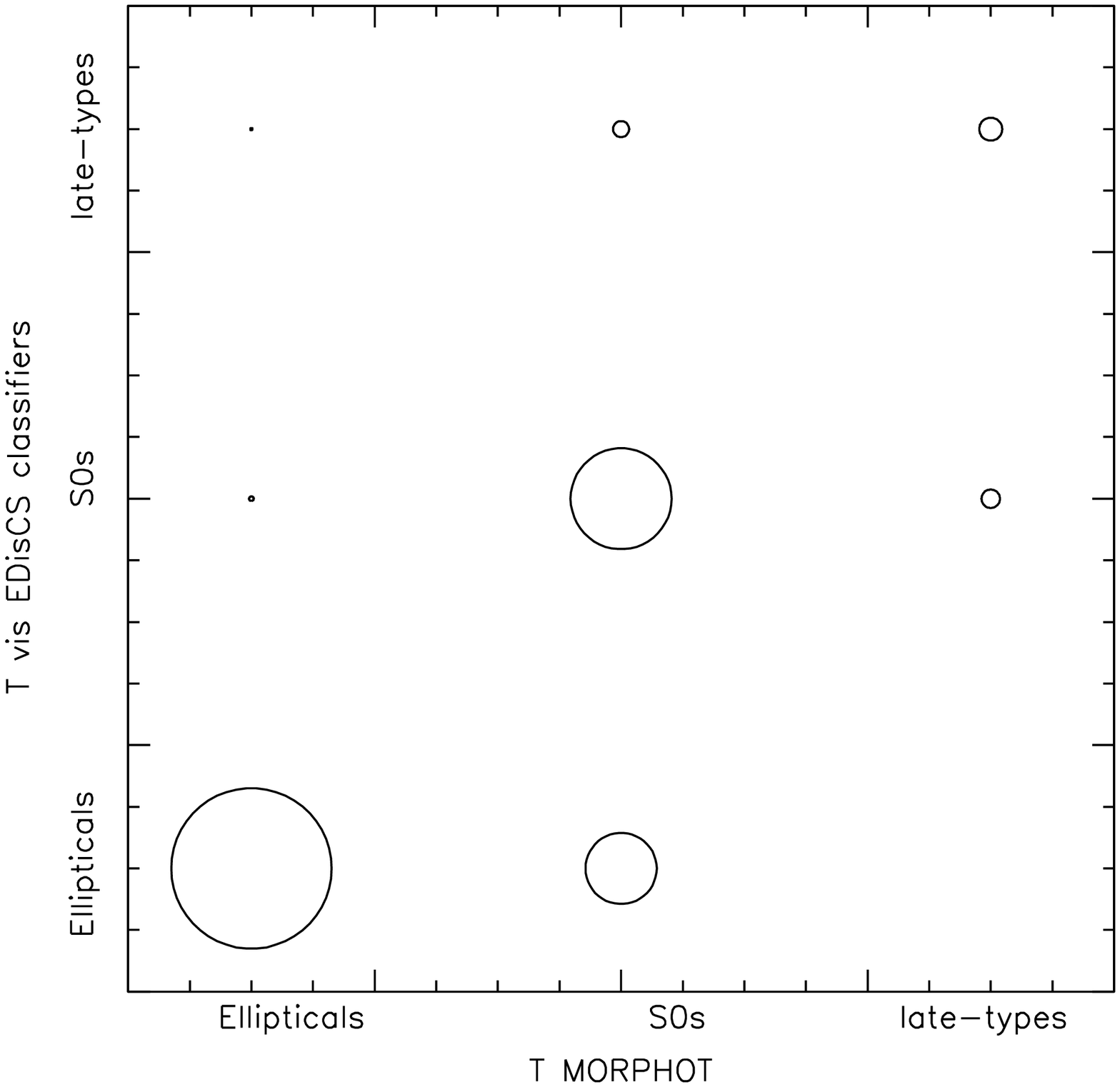}
\vspace*{2.2cm}
\end{minipage}
\hspace{1cm}
\caption{Comparison between the visual classification performed by the EDisCS classifiers
and the WINGS visual classification (left panel)  and the automatic classification performed by MORPHOT (right panel). The circle radius is proportional to the number of galaxies.\label{mor}}
\end{figure*}

Again, these discrepancies turn out to be similar
to the typical discrepancy among visual classifications given by
experienced, independent human classifiers, so we conclude that the
different methods adopted provide a comparable classification.

\subsubsection{Mass limited sample} \label{w_m}
The spectroscopic magnitude limit of the WINGS survey is
 V=20. Considering the distance module of $\sim 37.5$ of the most
 distant WINGS cluster, and a reddest color of $(B-V)=1.2$, the
 magnitude limit corresponds to a mass limit $M_{\ast} = 10^{9.8}
 M_{\odot}$, above which the sample is unbiased.  
Adopting this limit, the final sample composed by galaxy
 cluster members, within 0.6$R_{200}$, excluded the BCGs,
consists of 1233 galaxies, of which
396 ellipticals galaxies, 586 S0s, 239 late
 type galaxies and 12 galaxies with no classification (see Table~3). 
The corresponding numbers weighted 
for completeness
are also given in Table~3, for a total of 1894.42 galaxies.
Numbers of WINGS galaxies above the EDisCS mass limit 
($M_{\ast} = 10^{10.2} M_{\odot}$, see below) are also listed
in Table~3.
 
\subsubsection{Magnitude limited sample}
Since we will use the morphological fractions derived by \cite{poggianti09}, 
we use the magnitude limit adopted in that study,
which is M$_{V}$=-19.5, corresponding to the passively evolving
limit of the EDisCS morphological study at higher redshifts \citep{desai07}.
The final sample of members within 0.6$R_{200}$, excluded the BCGs, 
consists of 263 ellipticals, 410 S0s, 191 
late-type galaxies, and 10 galaxies with no morphological classification,
for a total of 874 galaxies, as listed in Table~3. 
Numbers weighted for completeness, for a total of 1340.73
galaxies, can be found in Table~3.

\subsection{High-z sample: EDisCS}\label{sample_ed}
For EDisCS, 
there are 605  spectroscopically confirmed members. 381 of them ($\sim 63\%$)
have HST morphologies. 
The spectroscopic magnitude limit ranges
between I=22 and I=23 depending on redshift, and the most conservative
spectroscopic mass limit using a \cite{kr01} IMF becomes $M =10^{10.6} M_{\odot}$
\citep{vulcani10}.  Therefore, relying only on spectroscopy, the lower
mass limit of our study would be very high, and numbers would be
relatively small.

For this reason, we decided to use all photo-z members (see 
\S \ref{dataediscs}).  All the EDisCS mass functions analyzed below
are based solely on the photo-z membership, unless otherwise stated.
The photo-z technique allows us to push the mass limit to much
lower values than the spectroscopy (see below). As shown in Appendix A,
spectroscopic and photo-z techniques give very consistent results
in the mass range in common. 

As for the WINGS sample, BCGs and galaxies at radii greater than $r=0.6R_{200}$
have been excluded from the analysis. 

Morphologies are
discussed in detail in \cite{desai07}.  The morphological
classification of galaxies is based on the visual classification of
{\it HST/ACS} F814W images sampling the rest-frame $\sim
4300-5500$ \AA{} range, similarly to WINGS.

\subsubsection{Mass limited sample} \label{e_m}
The magnitude completeness limit of the EDisCS photometry is $I \sim
24$, though the completeness clearly remains high to magnitudes significantly
fainter than $I = 24$ \citep{white05}.

We consider the most distant cluster, {\it
cl 1216.8-1201}, that is located at $z\sim0.8$ and determine the value of the
mass of a galaxy with an absolute B magnitude corresponding to $I=24$,
and a color $(B-V) \sim 0.9$, which is the reddest color of galaxies
in this cluster.
In this way, the EDisCS mass completeness limit based on photo-z 
is $M_{\ast} = 10^{10.2} M_{\odot}$.\footnote{Note that with this selection
only a few (1.7\%) galaxies in the mass-limited sample are fainter than I=23,
and therefore have no morphological classification from Desai et al. 2007.}

The
completeness is confirmed by the analysis of \cite{rudnick09}, who find that
for the magnitudes used in this paper ($M_{g}$ always $\leq -20.2$)
the photo-z counts are fully consistent with the statistically background
subtracted counts for both red and blue galaxies.

The final mass-limited EDisCS sample above this limit
consists of 489 galaxies, 156 of which are classified as ellipticals,
64 as S0s, 252 as late-type galaxies, and 17 galaxies with unknown
morphology (see \tab\ref{numb}).

\subsubsection{Magnitude limited sample}
We use the same 
magnitude limit as in the EDisCS morphological study \citep{desai07},
M$_{V}$=-20, which for passive evolution corresponds to the
magnitude limit in the WINGS morphological study \citep{poggianti09}.
With this limit, our sample consists of 544
galaxies, 151 of which are ellipticals, 67 S0s, 304 late-types, and 22
galaxies with unknown morphologies (see \tab\ref{numb}).

\begin{table*}
\centering
\begin{tabular}{|c||c|c|c|c||c|c||c|c|c}
\hline
& \multicolumn{6}{|c|}{WINGS} 	&& \multicolumn{2}{|c|}{EDisCS} \\ 
	& \multicolumn{2}{|c|}{$M_{\ast} \geq 10^{9.8} M_{\odot}$} & \multicolumn{2}{|c|}{$M_{\ast} \geq 10^{10.2} M_{\odot}$} & \multicolumn{2}{|c|}{M$_{V} \leq -19.5$}&& $M_{\ast} \geq 10^{10.2} M_{\odot}$ &M$_{V} \leq -20$\\
\hline
            	&N$_{obs}$	& N$_{w}$	&N$_{obs}$	& N$_{w}$ 	&N$_{obs}$	& N$_{w}$	&&N	&N	\\
\hline	
all	&1233		&1894.42		&737 		&1132.55		&874		&1340.73  	&&489	&544\\
ellipticals	&396		&603.31		&243		&367.12		&263		& 397.33		&&156	&151\\
S0s	&586		&912.45 		&369		&578.13		&410		&641.11		&&64	&67\\
late-types&239		&358.52 		&119		&176.84		&191 		&286.13		&&252	&304 \\
unknown	&12		&20.14 		&6		&10.46		&10		& 16.16		&&17	&22\\
\hline
\end{tabular}
\caption{Number of galaxies above the different limits. For WINGS both observed numbers and numbers weighted for spectroscopic incompleteness
are given \label{numb}}
\end{table*}

\section{Results: The mass function}
In this section we analyze only the mass limited samples of both
data-sets.\footnote{Appendix B presents the mass functions for
magnitude limited samples of WINGS and EDisCS galaxies, and clearly
shows the biases inherent in this type of selection.}

We use galaxy stellar masses to characterize the mass distribution of 
different galaxy types in clusters, both at low- and high-z. 

Above the completeness limit, we build histograms to define the mass
distribution. In each mass bin, we sum all galaxies of all clusters to
obtain the total number of galaxies, then we divide this number 
for the width of the bin, to have the number of
galaxies per unit of mass. The width of each mass bin is 0.1 dex. 
In the case of spectroscopic data, as for WINGS, in
building histograms each galaxy is weighted by its incompleteness
correction.
Errorbars are computed using poissonian errors \citep{gehrels86}.

Clusters of different masses could have a different
role in the total mass function, so we have tried to test the importance of 
this effect. For WINGS galaxies, we have determined the mass functions
separately for clusters with $\sigma <700$ $km \,s^{-1}$, 
$700$ $km \,s^{-1}<\sigma <800$ $km \,s^{-1}$ and $\sigma >800$ $km \,s^{-1}$
(plots not shown) and we found no differences. 
Similarly, for EDisCS galaxies
we have tried building the total mass function 
normalizing for the number of galaxies in each cluster, 
to avoid the possible  larger contribution of  larger clusters to the total
mass function. Once again, 
we have found  (plots not shown) no differences with the mass function presented in this paper. 
So, a different combination of the mass functions of different clusters 
could only influence the final normalization; anyway, since we are interested
mainly in the shape of the mass distributions and not in the counts (see below), this
doesn't affect our results. 
Consequently, we conclude that the mass of the cluster, over the
cluster mass ranges considered here, does not influence significantly the 
shape of the
total mass function, and in the following we use the simple sum of all
galaxies in all clusters.

To quantify the differences between different mass functions, we
perform Kolmogorov-Smirnov (K-S) tests. However, 
the standard
K-S test does not consider the completeness issues when it  builds the
cumulative distribution (since it assigns to each object a weight
equal to 1).  So we modified the test, to make the relative importance
of each galaxy in the cumulative distribution depend on its weight.
In the following, we will always use this
modified K-S test. Obviously, for photo-z data all galaxies have a
weight equal to 1, and using the modified test is equivalent to using 
the standard one.

\subsection{The  total mass function and its evolution}
\begin{figure}
\centering
\includegraphics[scale=0.45]{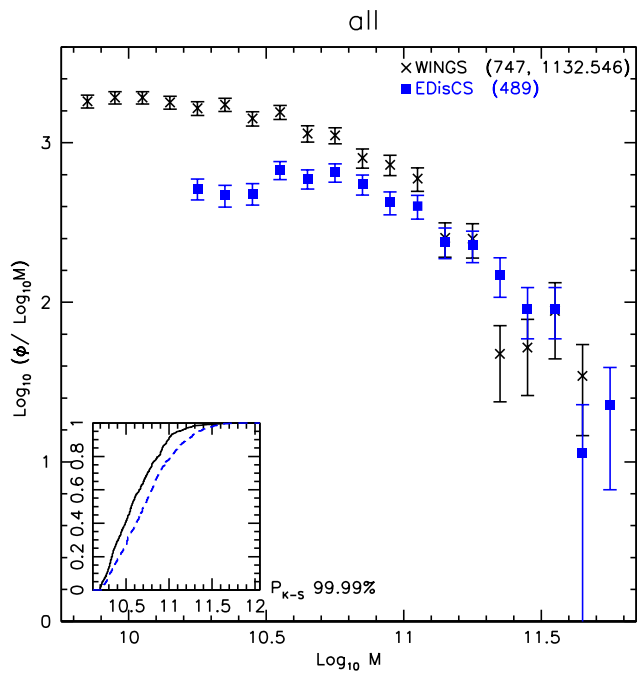}
\caption{Comparison of the mass distribution of EDisCS (blue filled
squares) and WINGS (black crosses) galaxies of all
morphological types, for the mass limited sample.  Mass distributions
are normalized to have the same total number of objects at $\log M_{\ast} / M_{\odot}
\geq 11 $. Errors are poissonian.  Only points above the
respective completeness limits are drawn.  Numbers in the labels are
the total number of galaxies above the highest of the two completeness
limits. For WINGS, both observed (first) and weighted (second) numbers
are given.  The inset shows the cumulative distributions, with the
relative K-S probability, for galaxies above the common mass limit. 
The mass distributions depend on redshift: for a similar number of 
higher mass galaxies, for $\log M_{\ast} / M_{\odot}
\leq 10.8$ there are proportionally less massive galaxies at low-z
than at high-z.  \label{all_norm}}
\end{figure}

We first compare the mass distribution of all 
galaxies at low and high redshift.
As we are interested in the shapes of the mass
 functions and not in the total number of galaxies, 
we normalize
the mass distribution of EDisCS galaxies to the mass distribution of
WINGS galaxies to have the same total number of galaxies
above $\log M_{\ast} / M_{\odot} \geq 11$.  In this way we assume
that the most
massive galaxies are already in place at z$\sim 1$ 
( see, e.g., \citealt{fontana04, pozzetti07}). 

In \fig\ref{all_norm} the mass distribution of WINGS galaxies (crosses) is
compared to the mass distribution of EDisCS galaxies (squares).  The comparison
is meaningful only above the highest of the two mass completeness
limits, that is for $\log M_{\ast} / M_{\odot} \geq 10.2$.

The two mass distributions are very different.  Indeed, the EDisCS
mass function is flat for $\log M_{\ast} / M_{\odot} \leq 10.8$.  The WINGS mass
function is rather flat up to about $ \log M_{\ast} / M_{\odot} \sim 10.6 $ and
then it begins to decline more steeply, with a massive end slope similar 
to that of the
EDisCS mass function.  The striking difference is the dearth of
galaxies less massive than $\log M_{\ast} / M_{\odot} \leq 10.8$ in EDisCS
compared to WINGS, for the same number of massive galaxies.  The K-S
test confirms that the two distributions are different at the 
99.99\% level. We remind  that the results of the K-S test do not depend
on the normalization we adopted in the plots. So, also using other normalization
criteria, we would reach the same conclusions.

We conclude that the total mass distribution of galaxies in clusters evolves
with redshift.  In a mass-limited sample, cluster galaxies at high-z
are on average more massive than in local clusters. Assuming no
evolution at the massive end, the population of $\log M_{\ast} / M_{\odot} \leq 10.8$
 galaxies must have grown significantly between $z=0.8$ and
$z=0$.

In principle, the evolution of the galaxy stellar mass function can be
due to several factors: a) strong environmental  mass segregation, for which galaxies
infalling into clusters between the two redshifts have a different
mass distribution than galaxies already in clusters at high-z
(infalling galaxies being on average
less massive); b) mass loss from massive
galaxies due to harassment; c) galaxy merging; d) mass growth due to
star formation.  In \S7 we will discuss which effects are likely to play an
important role and which ones can probably be neglected in this study.

\subsection{The mass functions of different galaxy types}
We now analyze separately the mass function of each morphological
type, to see if and how the distribution of masses depends on the galaxy type.
In this case, we do not apply any normalization among
the different mass functions, to show 
what morphological type dominates as a function of mass.
\begin{figure*}
\begin{minipage}[c]{175pt}
\centering
\includegraphics[scale=0.45,clip = false, trim = 0pt 150pt 0pt 0pt]
{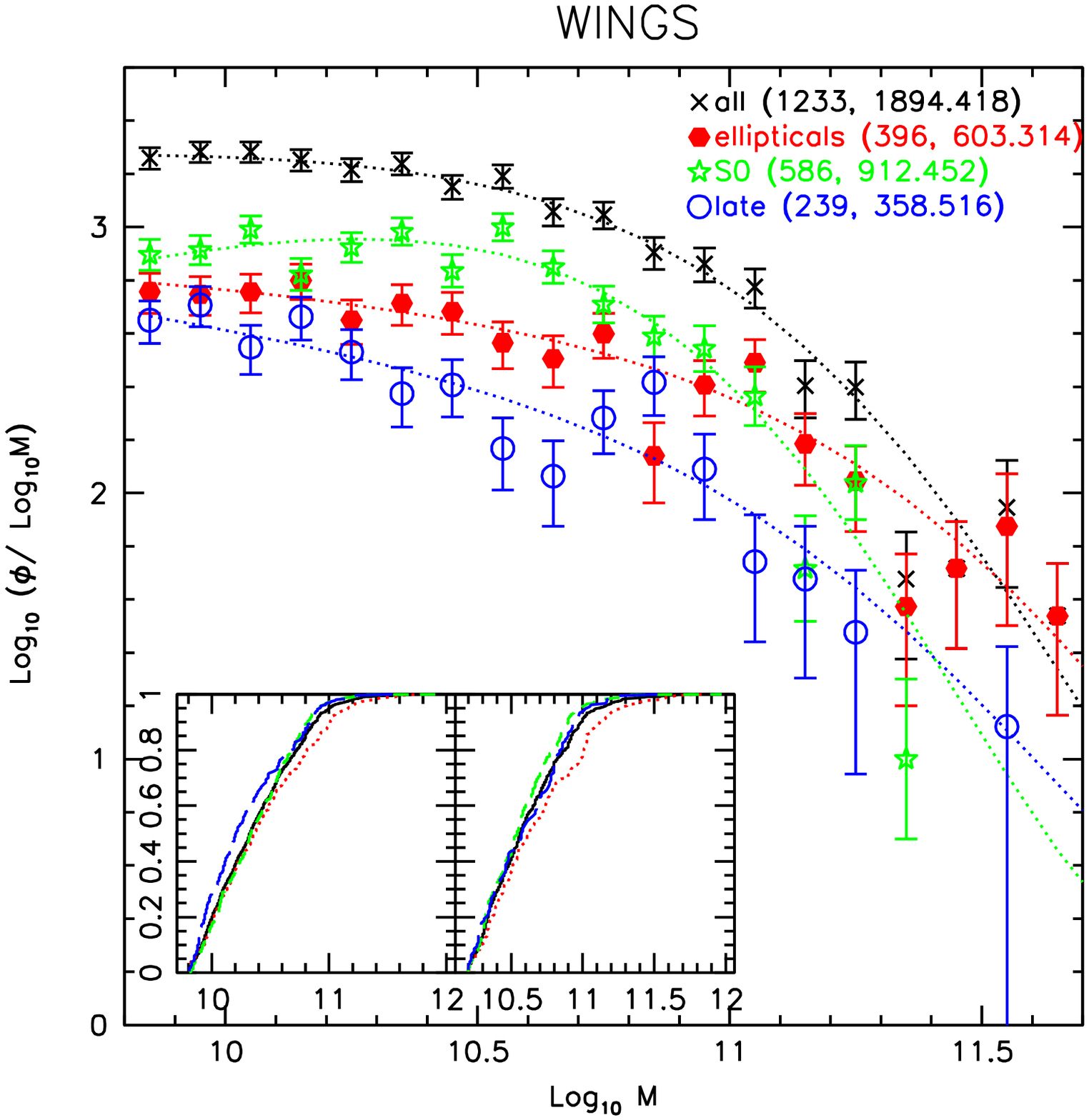}
\vspace*{2.2cm}
\end{minipage}
\hspace{2.5cm}
\begin{minipage}[c]{175pt}
\centering
\includegraphics[scale=0.45,clip = false, trim = 0pt 150pt 0pt 0pt]
{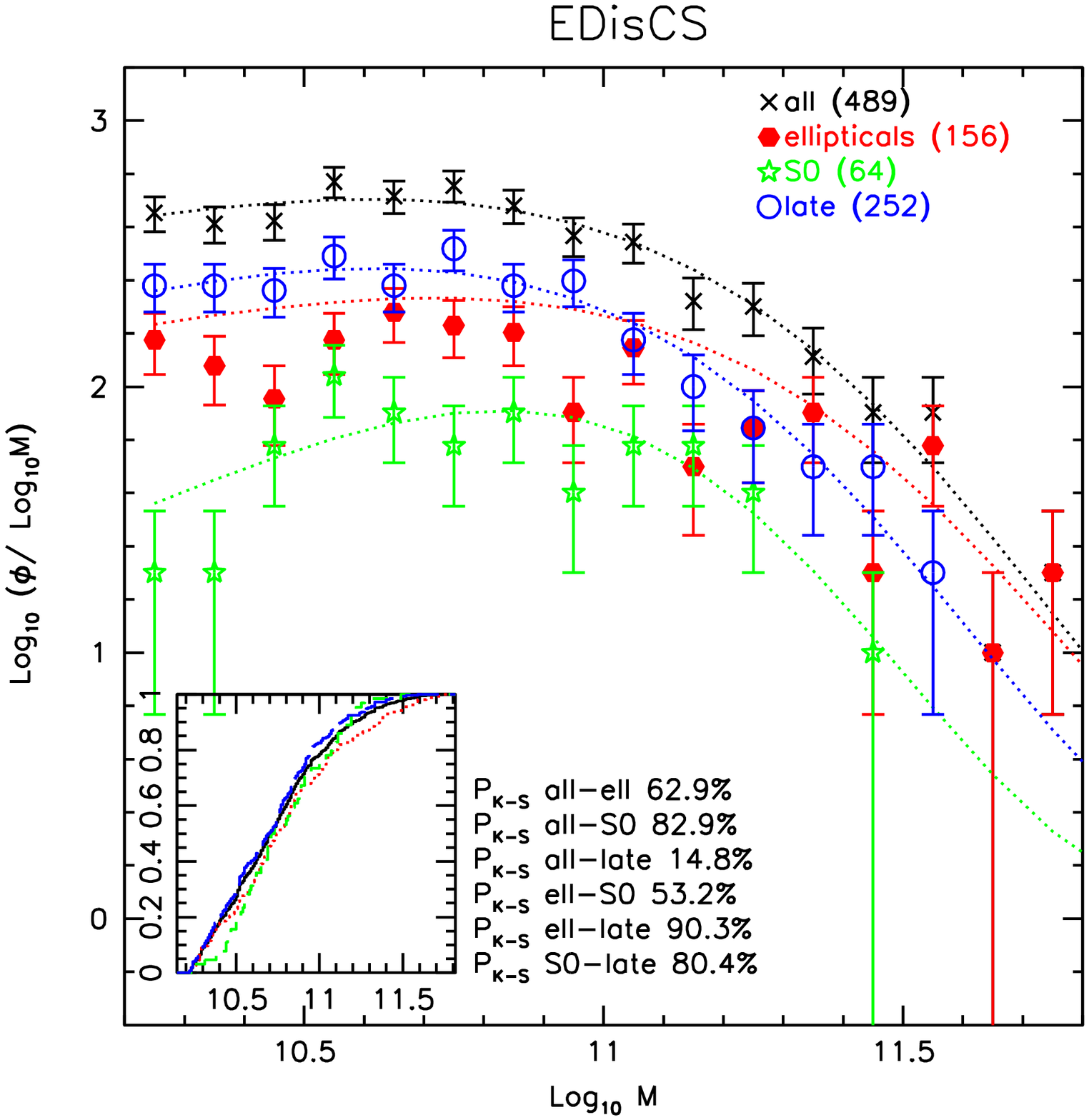}
\vspace*{2.2cm}
\end{minipage}
\hspace{1cm}
\caption{Mass distribution of WINGS galaxies (left panel) and 
EDisCS galaxies (right panel).  In the insets, the cumulative
distributions are shown. For WINGS, the right inset is done
for the EDisCS mass limit, $\log M_{\ast} / M_{\odot} =10.2$. Black
crosses represent all galaxies, filled red circles elliptical
galaxies, green stars S0s and blue open circles late-type galaxies.
Errors are defined as poissonian errors. Numbers in the labels are the
number of galaxies in each morphological class, above the respective
mass limit. For WINGS, both observed (first) and weighted (second) 
numbers are given. For EDisCS, the K-S probabilities are also given.
 Best-fit \citet{schechter76} mass function are showed dotted. The values of the parameters of the fit
are presented in \tab\ref{sch}. At low-z, different morphological types have different mass functions, while
at high-z no conclusion can be drawn.\label{massf}}
\end{figure*}
\fig\ref{massf} shows the mass distributions for galaxies of the WINGS
dataset (left panel) and of the EDisCS dataset (right panel). 
Galaxies of different morphological types are indicated by different
colors and symbols. 

In \tab\ref{sch} we give the analytic expression for the mass functions of the different morphological
types for both samples, estimating
 the best-fit \cite{schechter76} parameters ($\alpha$, $M^{\ast}$, $\phi^{\ast}$).

\begin{table*}
\centering
\begin{tabular}{|c||l|c|c|c||c|}
\hline
	&	&	$\log M^{\ast}$ & 	$\alpha$ &$\phi^{\ast}$\\
\hline	
WINGS &	all galaxies	&11.667$\pm$0.052 & 	-0.987$\pm$0.009 &	1.552$\pm$0.055\\
	&	early-types 	&11.643$\pm$0.055&	-0.977$\pm$0.011&	1.526$\pm$0.059\\
	&	ellipticals		&11.879$\pm$0.076&	-1.017$\pm$0.014&	1.13$\pm$0.067\\
	&	S0s			&11.391$\pm$0.052&	-0.925$\pm$0.020&      1.680$\pm$0.089\\
	&	late-types		&11.683$\pm$0.171&	-1.04$\pm$0.031&	0.992$\pm$0.138\\
\hline
EDisCS &   all galaxies	&11.684$\pm$0.061 &	-0.915$\pm$0.026 &1.577$\pm$0.105	\\
	&	early-types	& 11.700$\pm$0.081 &	-0.896$\pm$0.041 &1.423$\pm$0.138 \\
	&	ellipticals		& 11.847$\pm$0.126 &	-0.947$\pm$0.044& 1.159$\pm$0.148 \\
	&  	S0s			&11.300$\pm$0.125 & 	-0.662$\pm$0.147 &1.679$\pm$0.235\\
	&	late-types		&11.523$\pm$0.067 & 	-0.875$\pm$0.044 	&1.560$\pm$0.141 \\
	&	&&&&\\
\hline
\end{tabular}
\caption{Best-fit \citet{schechter76} parameters ($\alpha$, $M^{\ast}$, $\phi^{\ast}$) for the galaxy mass functions of the different morphological types discussed in this paper. Fits are perfomed directly on the logarithmic points.\label{sch}}
\end{table*}

For WINGS, the shape of the mass distribution strongly depends on morphological
type.  
The total mass function is dominated by S0 galaxies up to
$\log M_{\ast}/M_{\odot} \sim 11$, and by elliptical galaxies at higher
masses. The total mass function is quite flat up to $\log M_{\ast} / M_{\odot} \sim 10.6$, 
as it is the function of both S0s and ellipticals.
Instead, the mass function of late-type galaxies declines over the whole
mass range. 
While there are elliptical galaxies at all masses, there are
no  S0s or  late-types more massive than $\log M_{\ast} / M_{\odot}\sim11.4$ and
$\log M_{\ast} / M_{\odot}\sim11.6$ respectively.

We perform a K-S test, both for a mass limit
$\log M_{\ast} / M_{\odot} \geq 9.8$ (the WINGS limit), and for $\log M_{\ast} / M_{\odot} \geq
10.2$ (the same limit as EDisCS).
The K-S results are
summarized in \tab\ref{K-S}.\footnote{We note that with the term
``all'' we refer to all galaxies regardless of their morphological
type and we are including also galaxies for which we do not have a
morphological classification, that are only a few (see Table~3).
However, we have checked that the mass distribution is the same
considering all galaxies or only the sum of ellipticals, S0s and
late-type galaxies.}

\begin{table*}
\centering
\begin{tabular}{|l|c|c||c|c|}
& \multicolumn{2}{c}{WINGS} & &EDisCS\\
\hline
 & $P_{K-S}$ ( $M_{\ast} \geq 10^{9.8} M_{\odot}$) & $P_{K-S}$ ( $M_{\ast} \geq 10^{10.2} M_{\odot}$) && $P_{K-S}$ ( $M_{\ast} \geq 10^{10.2}M_{\odot}$)\\
\hline
ell-S0 & 99.84\%&  99.99\% && 53.18\%\\
ell-late & 99.44\%&  98.20\%&&90.20\%\\
S0-late & 99.72\%&  91.40\%&&80.43\%\\
\hline
all-ell &  99.38\%&  99.9\%& & 62.98\%\%\\
all-S0 &  69.52\%&  95.53\%&& 82.89\%\\
all-late & 98.64\%&  24.94\%& & 14.77\%\\
\hline
\end{tabular}
\caption{Results of the Kolmogorov-Smirnov test performed on the mass
function of different morphological types for the WINGS and EDisCS mass limited
sample. For WINGS, results also above the EDisCS mass completeness
limit are given. $P_{K-S}$ is the probability that two distributions are drawn
from different parent distributions.  The K-S has been modified to
take into account the different weights of galaxies observed
spectroscopically. \label{K-S}}
\end{table*}

For $\log M_{\ast} / M_{\odot} \geq 9.8$, the morphological mass functions
significantly differ one from another: ellipticals, S0 and late-type
galaxies all have different mass functions, with the K-S that excludes
the possibility that they are drawn from the same parent distribution
with a probability always greater than 99.4\%.  For $\log M_{\ast} / M_{\odot} \geq 10.2
$, S0s and late-type galaxies are not statistically
distinguishable any more (probability = 91.4\%). 
This suggests that, except for the least
massive galaxies, the S0 and spiral mass functions could in fact be
similar. This could be due to the fact that when we 
analyze the sample at $\log M_{\ast} / M_{\odot} \geq 10.2$ we are actually
excluding Sc and Irregulars galaxies, that mostly have masses below this limit. 
However, in our sample, they represent only 5\% of the total population
of late-type galaxies, so we do not expect they matter a lot.

Comparing each type with all galaxies, for $\log M_{\ast} / M_{\odot} \geq 9.8$
only S0 galaxies could have a similar distribution to the sum of all
galaxies ($P_{K-S} \sim 70\%$),
while in the other two cases
the test indicates that the distributions are different, with a
probability at least $>98.5\%$.
Instead, above $\log M_{\ast} / M_{\odot} = 10.2$ 
the K-S test is less conclusive: it rejects the hypothesis of
similarity only between ellipticals with
all galaxies, with a probability of $99.9\%$.
This demonstrates how important it is to reach deep mass limits: in fact
the greatest differences among the mass functions tend to
emerge at lower masses.

For EDisCS (right panel in \fig\ref{massf}), 
the total mass function is dominated by the
mass function of late-type galaxies up to $\log M_{\ast} / M_{\odot} < 11 $
(and not by S0s galaxies as in WINGS).  
The most massive bins of the total function are dominated by
elliptical galaxies.  In fact, there are no S0s or late-types more
massive than  $\log M_{\ast} / M_{\odot}\sim11.5$ and  $\log M_{\ast} / M_{\odot}\sim11.6$
respectively.  The mass function of S0s is very noisy, due to
their low number  in the sample.

At high-z, the data cannot rule out that the mass functions of
the different morphological types are all similar, with K-S probabilities
always smaller than or at most equal to
 90\% (see \tab\ref{K-S}). We also split late-type galaxies
in Sa+Sb and Sc+Irr galaxies, to investigate if the two populations
behave differently (\fig\ref{ed_late}). 
Even though the K-S  seems to be not conclusive 
(it gives a probability of $\sim$ 93\%),
we find that mass functions are quite different (\fig\ref{ed_late}), indicating
that there are proportionally more lower mass galaxies ($\log M_{\ast} / M_{\odot} < 11 $) 
among Sc+Irr than among Sa+Sb.\footnote{We note that such an analysis cannot
be performed in WINGS, due to the very small number of Sc+Irr.}

\begin{figure}
\centering
\includegraphics[scale=0.45]{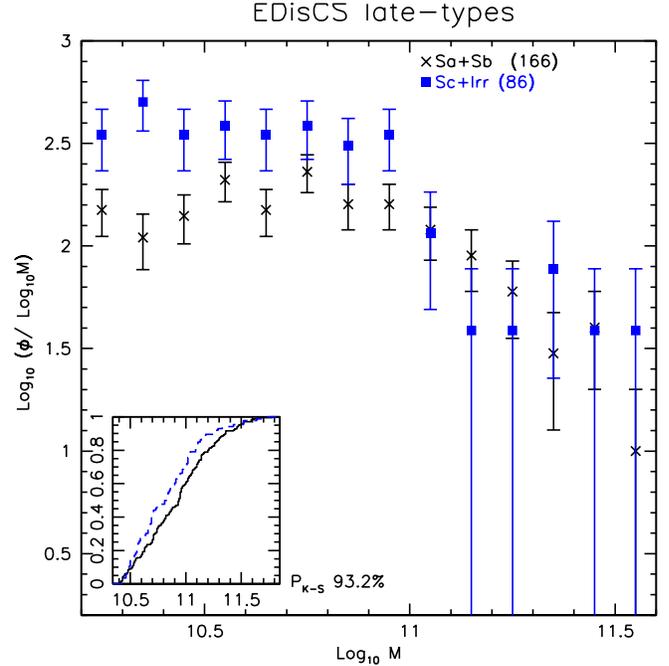}
\caption{Mass distribution of EDisCS late galaxies split into Sa+Sb (black crosses) 
and Sc+Irr (blue filled
squares), for the mass limited sample.  Mass distributions
are normalized to have the same total number of objects at $\log M_{\ast} / M_{\odot}
\geq 11$. Errors are poissonian.  Numbers in the labels are
the total number of galaxies above the completeness
limits. The inset shows the cumulative distributions, with the
relative K-S probability, above the common mass completeness limit. The two distributions
are quite different, indicating that early- and late-spirals have different mass distributions.   \label{ed_late}}
\end{figure}

Also the comparison between each morphological type
and all galaxies does not give conclusive results, because the  K-S probabilities
are always smaller than 90\%.
In WINGS, we have seen that, for the same mass limit, only S0s and late-types
could have the same mass distribution, while they both differ
from ellipticals. The lack of significance
of the differences in EDisCS might be due to poorer number statistics,
as well as to the higher mass limit. 

To check if the lack of significance in EDisCS is real or above all
due to poor statistics, 
we performed 100 Monte
Carlo simulations extracting randomly from the WINGS sample at 
$\log M_{\ast} / M_{\odot} \geq 10.2 $ a subsample with the same number of galaxies as
EDisCS (that is 156 ellipticals, 64 S0s and 252 spirals).  We found
that
the probability of the K-S test to be conclusive is very low in all
cases (at most in 56\% of the simulations when we compare ellipticals
and late-types). These results are consistent with the hypothesis 
that in EDisCS the lack of a significant difference between the mass functions
of different types could be due to poor number statistics.

The conclusion we can draw from this section is that, at low redshift,
different morphological types have significantly different mass
functions, although S0s and late-type galaxies more massive than $\log
M_{\ast}/M_{\odot} = 10.2$ may have similar mass distributions.  
No conclusion can be drawn at high redshift.

\subsection{Evolution of the mass functions of different morphological types}
We now directly compare the mass distribution of each morphological
type in the two reshift bins, with the aim to understand if the shape
of the mass function of each given type evolves with
redshift, and what type of galaxies drives the evolution of the total
mass function.  As for the total mass function, we normalize the mass
distribution of EDisCS galaxies to the mass distribution of WINGS
galaxies to have the same total number above $\log M_{\ast} / M_{\odot} \geq
11$, since we want to compare the shape of the distributions.

\begin{figure*}
\begin{minipage}[c]{175pt}
\centering
\includegraphics[scale=0.45,clip = false, trim = 0pt 150pt 0pt 0pt]
{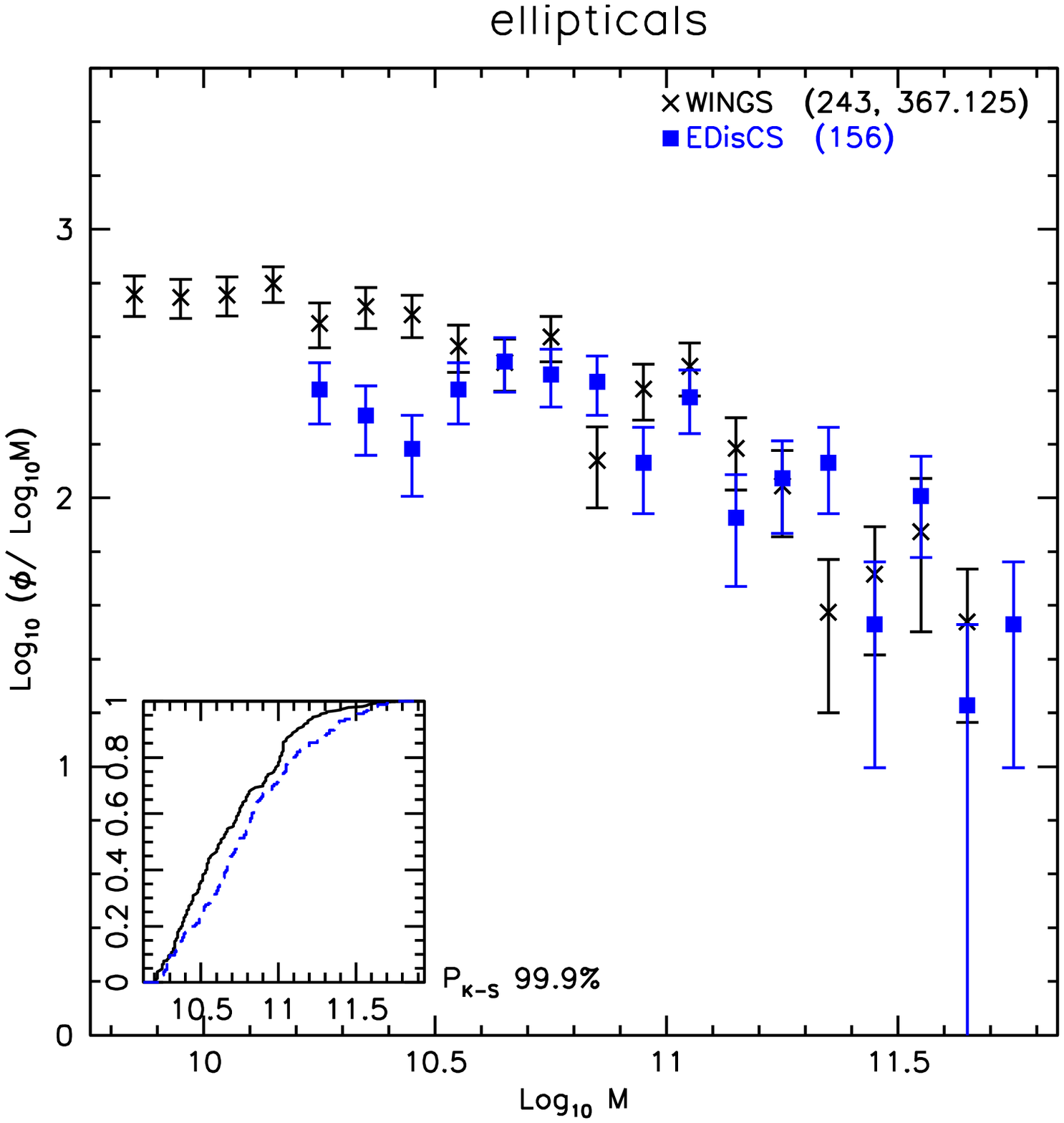}
\vspace*{2.2cm}
\end{minipage}
\hspace{2.5cm}
\begin{minipage}[c]{175pt}
\centering
\includegraphics[scale=0.45,clip = false, trim = 0pt 150pt 0pt 0pt]
{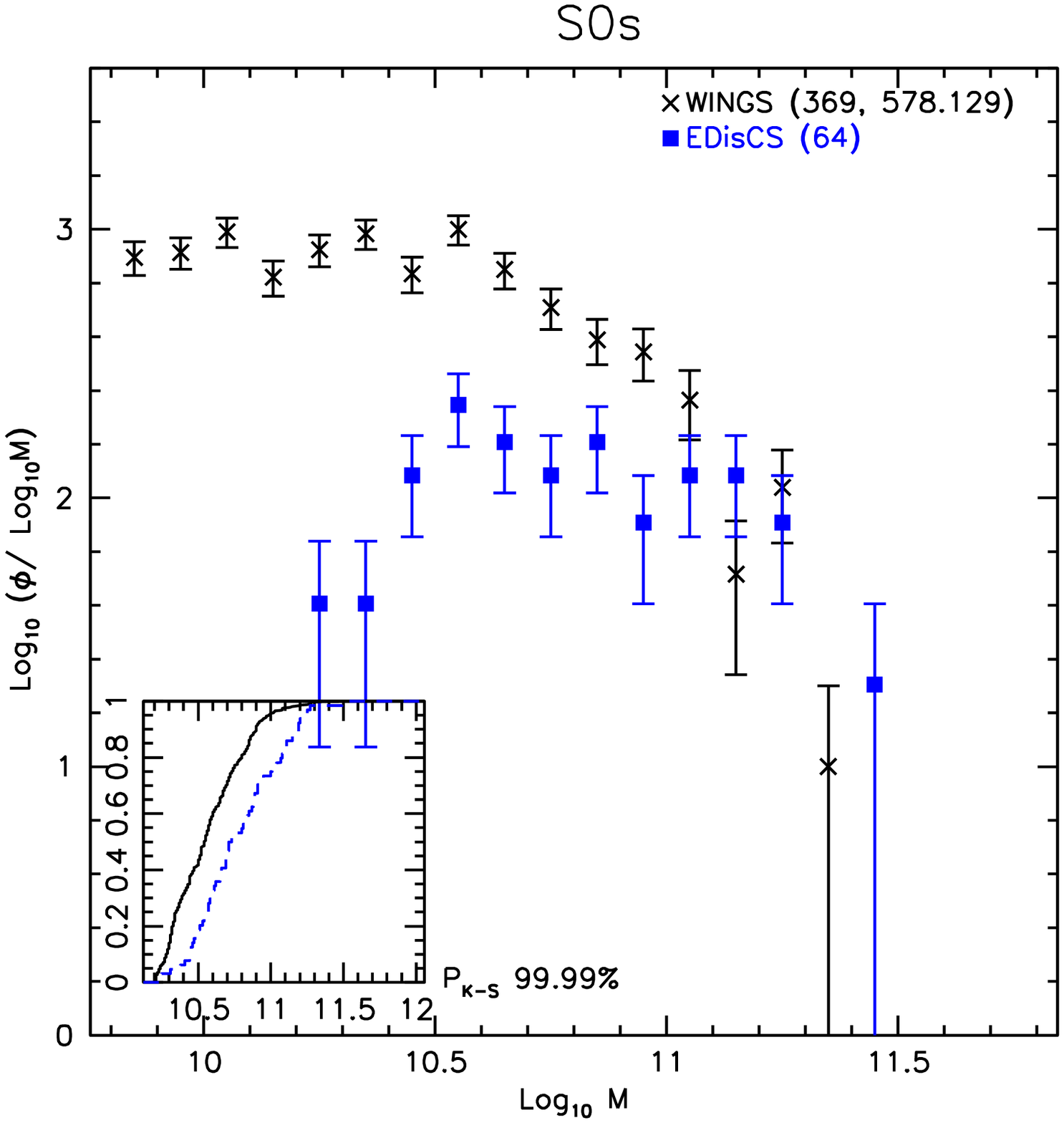}
\vspace*{2.2cm}
\end{minipage}
\hspace{2.5cm}
\begin{minipage}[c]{175pt}
\centering
\includegraphics[scale=0.45,clip = false, trim = 0pt 150pt 0pt 0pt]
{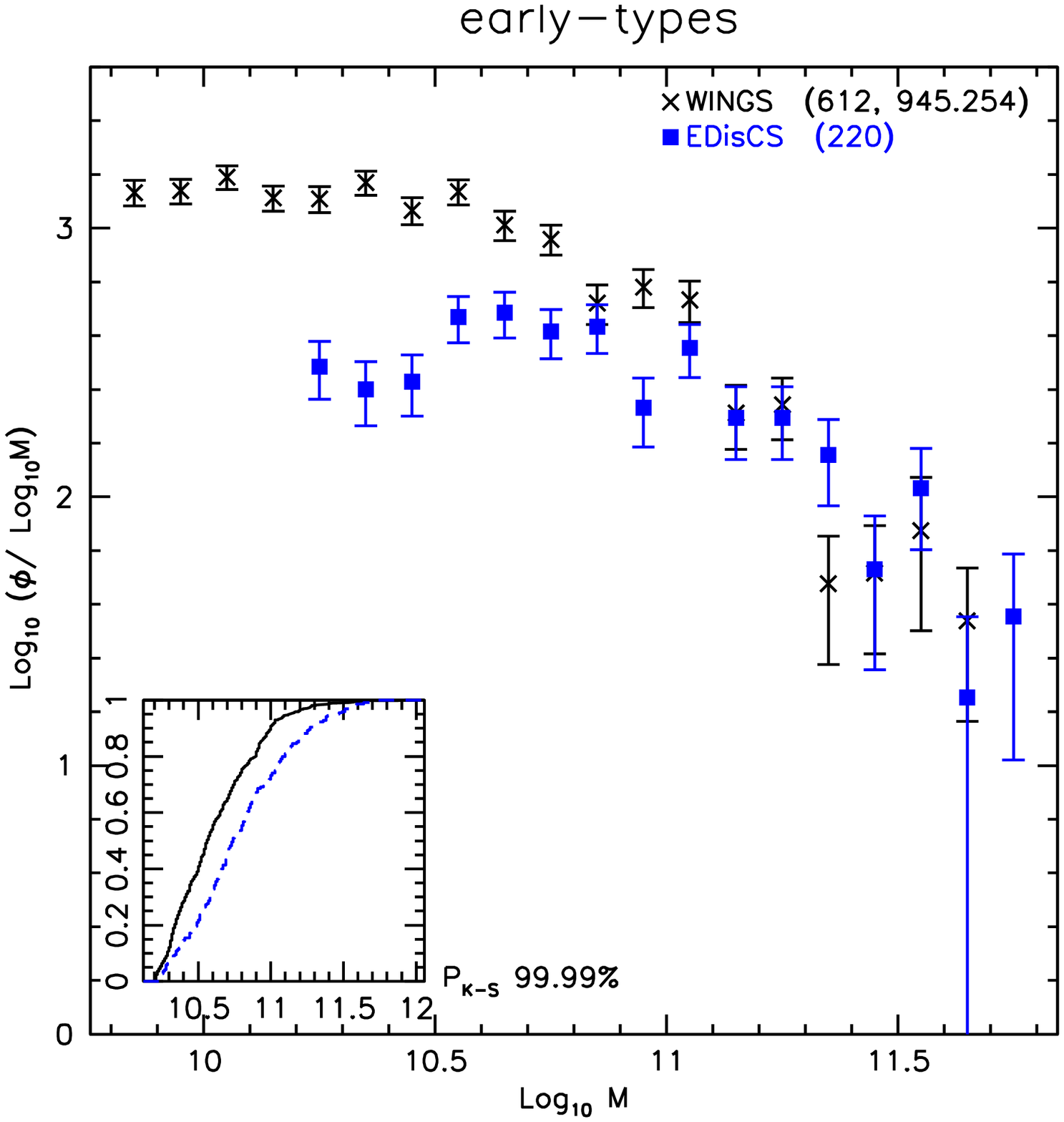}
\vspace*{2.2cm}
\end{minipage}
\hspace{2.5cm}
\begin{minipage}[c]{175pt}
\centering
\includegraphics[scale=0.45,clip = false, trim = 0pt 150pt 0pt 0pt]
{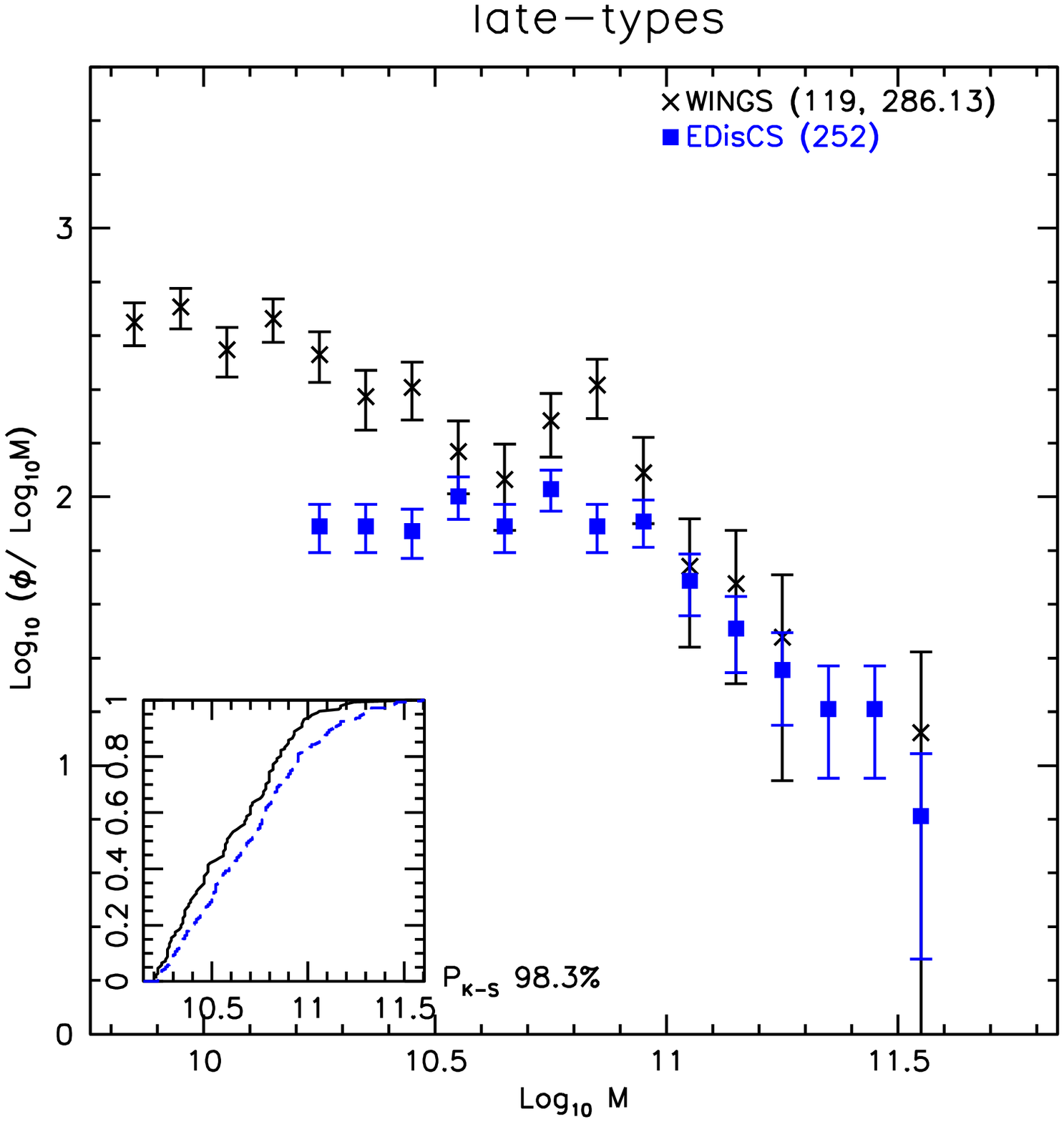}
\vspace*{2.2cm}
\end{minipage}
\hspace{1cm}
\caption{Comparison of the mass distribution of EDisCS (blue filled
squares) and WINGS (black crosses) galaxies, for the different
morphological types. Mass distributions are normalized to the WINGS total
number of objects with $\log M_{\ast} / M_{\odot} \geq 11$. Errors are
defined as poissonian errors. Numbers in the labels are the number of
galaxies in each morphological class, above the respective mass
limit. For WINGS, both observed (first) and weighted (second) 
numbers are given. The cumulative
distributions are shown in the insets. Each morphological type has a mass distribution
that depends on redshift. In WINGS there are always proportionally more less-massive
 galaxies than in EDisCS.\label{types}}
\end{figure*}

\fig\ref{types} compares the mass distribution of galaxies of each
morphological type at high and low redshift. 
The mass distribution of ellipticals is shown in the top left panel.
For both WINGS and EDisCS, the mass distribution starts rather flat at 
low masses, with a hint for a possible dip in EDisCS,
then begins to steepen at higher masses.
Statistically, the EDisCS and WINGS distributions are significantly
different.
As for the total mass function, there is a deficit of less massive
galaxies ($\log M_{\ast} / M_{\odot} \leq 10.6 $)
at high-z compared to low-z, for a given number of more massive galaxies. 
A K-S test rejects the null hypothesis of similarity of the two
distributions with a probability $>$99.9\%.

In the top right panel of \fig\ref{types} the mass distribution of S0
galaxies is shown.  It changes dramatically with redshift. WINGS
galaxies have a flat mass function up to $\log M_{\ast} / M_{\odot} \sim10.6$ and then
this becomes steeper. Instead, the mass function of EDisCS galaxies shows
a rise, a peak at about $\log M_{\ast} / M_{\odot} \sim 10.6 $, and a fall.
The K-S test states that the mass distributions at the two redshifts are
driven from two different parent distributions with a probability
$>99.99\%$. This result emerges with a high statistical significance
even though S0 galaxies are very rare in the EDisCS sample.

In the bottom left
panel of \fig\ref{types} we analyze the mass function of early-type
 galaxies. This can be directly compared with the mass
function of late-types (bottom right panel).  
The differences existing in the total mass
distribution still remain evident in the mass function of early-type
galaxies. While the mass distribution of high mass galaxies is
rather similar at the two redshifts, the dearth of galaxies with $\log
M_{\ast}/M_{\odot} \leq 10.9$ in EDisCS is even more evident
than in the total mass function.  Once again, the K-S test 
rejects the possibility that the two mass functions derive from the same
distribution with a probability $>$99.9\%. Of course, this behavior
was expected, being the early-type galaxies simply the sum
of ellitpicals and S0s.

The bottom right panel of \fig\ref{types} shows the mass distribution
of late-type galaxies.  The shape of the mass distribution of EDisCS
and WINGS is different up to $\log M_{\ast} / M_{\odot} \sim 11 $ and then it
is rather similar at higher masses. For WINGS late-types, the mass
distribution is immediately declining, while that of EDisCS starts
flat.  The K-S test gives a probability of $\sim 98\%$ that the two
distributions are not driven from the same parent distribution.

The conclusion we can draw from this section is that the mass
distribution of {\it each} morphological type evolves with redshift. All
types have proportionally more massive galaxies at high- than at
low-z.

In addition to the processes mentioned in \S5.1 (infall, harassment,
merging and star formation) that can change the total
mass function, there is an additional effect that could
influence the observed evolution of the mass functions of each
morphological type: the morphological transformation from one type to
the other.

\section{Results: The morphological fractions}

\subsection{The Morphology-Mass relation}

\begin{figure*}
\begin{minipage}[c]{175pt}
\centering
\includegraphics[scale=0.45,clip = false, trim = 0pt 150pt 0pt 0pt]
{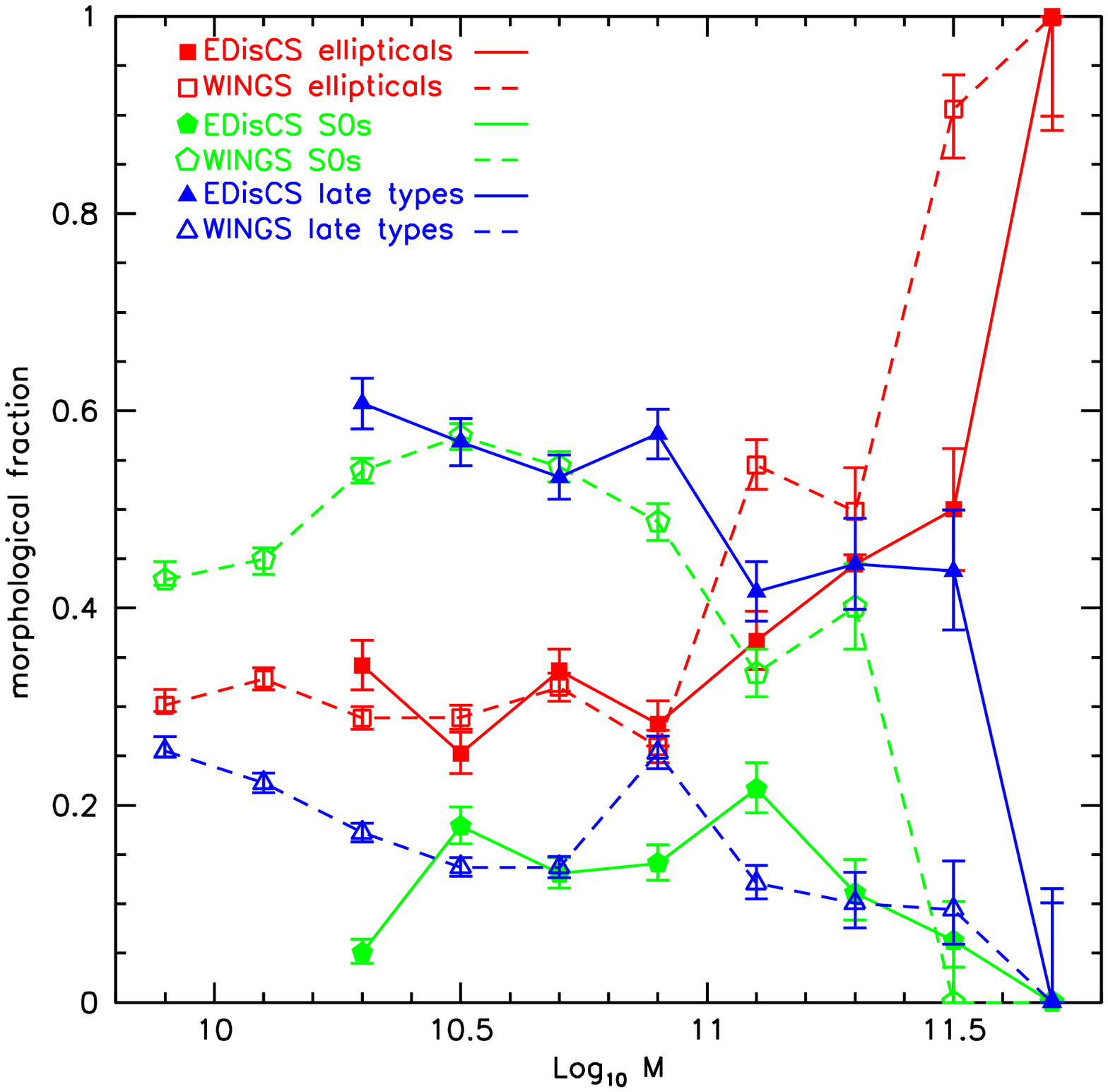} 
\vspace*{2.2cm}
\end{minipage}
\hspace{2.5cm}
\begin{minipage}[c]{175pt}
\centering
\includegraphics[scale=0.45,clip = false, trim = 0pt 150pt 0pt 0pt]
{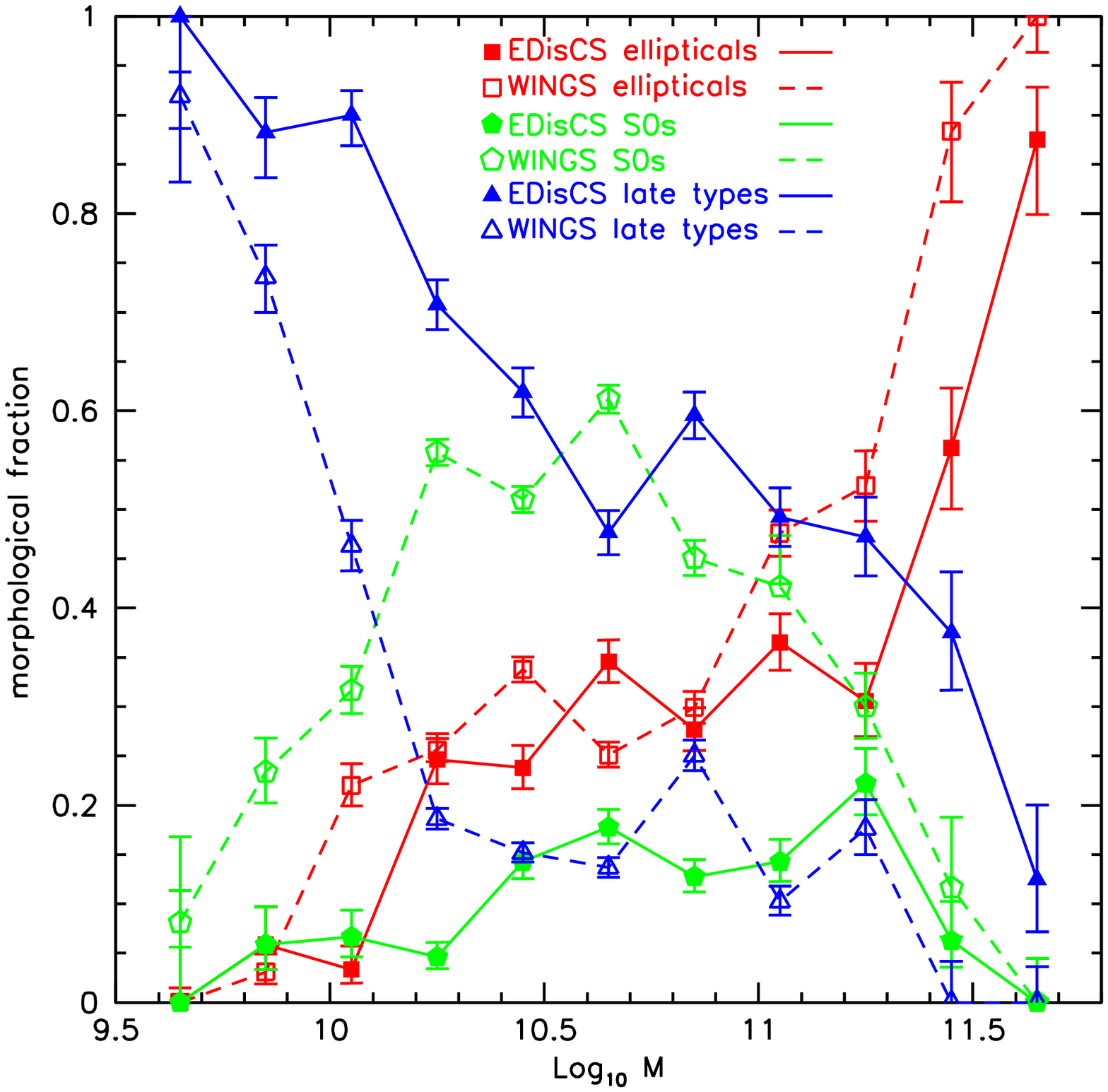} 
\vspace*{2.2cm}
\end{minipage}
\hspace{1cm}
\caption{Morphological fractions for the mass limited (left panel) and
magnitude limited (right panel) samples for galaxies at different
redshifts. Solid symbols and lines refer to EDisCS, empty symbols
and dotted lines refer to WINGS. Squares refer to ellipticals, 
circles refer to S0s,
triangles refer to late-types. Errors are defined as 
binomial errors. In both samples, the morphological mix changes with
galaxy stellar mass and with redshift at a given mass (see text for details). \label{frac}}
\end{figure*}

It is interesting to analyze how the incidence of each 
morphological type (the number of galaxies of that type over
the total) changes as a function of the
stellar mass (hereafter the morphology-mass relation, MMR). 

As in the following we will compare our findings with the results found by
\cite{desai07} and \cite{poggianti09} that use magnitude limited
samples, we discuss the MMR both in mass limited
and in magnitude limited samples.  Since above the completeness limit
of our samples the fraction of objects without morphological
classification is negligible (at most 4\%) in this analysis we do not
consider them, so we refer to ``all'' galaxies simply as the sum of
early and late-type galaxies.

\fig\ref{frac} shows the fractions of the different
morphological types as a function of galaxy mass, for both EDisCS and WINGS.  
In the left panel, the mass
limited samples are presented.  Errors are defined as binomial errors \citep{gehrels86}.

The trend of ellipticals at high and low redshift is similar: the elliptical
fraction is flat at low masses, then it increases at higher masses.
Ellipticals represent more than 50\% of galaxies at any mass greater than 
$\log M_{\ast} / M_{\odot} \sim 11$ at low-z and $\log M_{\ast} / M_{\odot} \sim 11.4 $ at
high-z. Thus, WINGS elliptical galaxies start to dominate at
a lower mass than in EDisCS.  The most massive galaxies tend to be
almost all ellipticals, at both redshifts.

The morphology-mass relation of S0s and late-types galaxies depends even
more strongly on redshift. In the Local Universe, for masses up to
$\log M_{\ast} / M_{\odot} \sim 11 $ S0s are by far the most common galaxies,
and their fraction is rather constant with mass up to $\log M_{\ast} / M_{\odot} \sim
10.9 $, 
whereas it decreases
steeply towards higher masses.  In contrast with the local Universe
where S0s are much more frequent at low than at high masses, at high-z
S0s are equally rare ($\leq 20\%$) at all masses.

Late-type galaxies are the
least represented morphological class in the Local 
Universe ($\leq 20\%$ at all masses), their
trend hardly depending on mass (excluding a possible fluctuation at $\log M_{\ast} / M_{\odot} \sim
10.9 $). 
At higher redshift, the situation is different: late-types are the most
common type of galaxies 
up to $\log M_{\ast} / M_{\odot} \sim 11.3$, with no clear trend 
with mass at lower masses, and a steep decline at higher masses.

In magnitude limited samples, 
the dependence of
the morphological fractions on mass is even more pronounced, at both redshifts.
At low masses,
late-type galaxies dominate the total population, while the presence
of elliptical galaxies is negligible.  On the contrary, at higher
masses elliptical galaxies are the most numerous, and late-type
galaxies the least numerous type of galaxies.  As far as S0s are concerned, 
while in WINGS they dominate at
intermediate masses ($\log M_{\ast} / M_{\odot} \sim 10.1-10.9$),
in EDisCS they do not dominate at any mass. 
Comparing high and low redshift, in general
there are many more intermediate-mass and
massive EDisCS late-types than WINGS late-types, that
instead dominate only the first bins up to $\log M_{\ast} / M_{\odot} \sim 10.1 $, and then fall to
zero by $\log M_{\ast} / M_{\odot} \sim 11.5$. Similarly to the mass-selected sample, WINGS ellipticals 
start to dominate at lower masses than in EDisCS.

These results show how the proportion of galaxies of different
morphological types changes with galaxy mass.  This MMR
 is significantly different in our mass-limited and
magnitude-limited samples.  In the mass-limited sample, the incidence
of each type {\it does not depend on mass} between our completeness mass
limits and $\log M_{\ast} / M_{\odot} \sim 11-11.1$ (flat trend), while it
strongly changes at higher masses. 
Both for the mass-limited and the
magnitude-limited samples, the MMR evolves
strongly with redshift, in connection to the evolution of the
morphological fractions described below.

\subsection{Evolution of morphological fractions}
\begin{table*}
\centering
\begin{tabular}{|c||c|c|c|c||c|c||c|c|c}
\hline
& \multicolumn{6}{|c|}{WINGS} 	&& \multicolumn{2}{|c|}{EDisCS} \\ 
	& \multicolumn{2}{|c|}{$M_{\ast} \geq 10^{9.8} M_{\odot}$} & \multicolumn{2}{|c|}{$M_{\ast} \geq 10^{10.2} M_{\odot}$} & \multicolumn{2}{|c|}{M$_{V} \leq -19.5$}&& $M_{\ast} \geq 10^{10.2} M_{\odot}$ &M$_{V} \leq -20$\\
\hline
            	&\%$_{obs}$         & \%$_{w}$	      &\%$_{obs}$	& \%$_{w}$ 	&\%$_{obs}$	& \%$_{w}$	&&\%	&\%	\\
\hline	
ellipticals	&32.4$\pm$1.3\% & 32.2$\pm$1.1\%  &	33.2$\pm$1.7\% 	&32.7$\pm$1.4\%	&30.4$\pm$1.6\%&30.0$\pm$1.3\%	&&33.0$\pm$2.2\%	&28.9$\pm$2.0\%	\\
S0s	&48.0$\pm$1.4\%    & 48.7$\pm$1.2\%  &	50.5$\pm$1.9\%	&51.5$\pm$1.5\%	&47.4$\pm$1.7\%	&48.4$\pm$1.4\%	&&13.6$\pm$1.6\%	&12.8$\pm$1.5\%	\\
early	&80.4$\pm$1.1\% &80.9$\pm$0.9\%    &83.7$\pm$ 1.4\%	&84.2$\pm$1.1\%	&77.8$\pm$1.4\%	&78.4$\pm$1.1\%	&&46.6$\pm$2.3\%	&41.8$\pm$2.1\%	\\
late-types&19.6$\pm$1.1\%&19.1$\pm$0.9\%  &	16.3$\pm$1.4\%	&15.8$\pm$1.1\%	&22.2$\pm$1.4\%	&21.6$\pm$1.1\%	&&53.4$\pm$2.3\%	&58.2$\pm$2.1\%	\\
\hline
\end{tabular}
\caption{Morphological fractions of galaxies in both mass and
mag-limited samples. Errors are computed as binomial errors. For WINGS, both observed and
completeness-weighted numbers are listed. Since galaxies with unknown
morphology are always $\leq 4\%$, they are not included in the
computation of fractions.\label{frac_morf}}
\end{table*}

The morphological fractions for both mass and magnitude limited samples
are given in \tab\ref{frac_morf}.  
WINGS fractions are given also above the EDisCS mass limit.
For WINGS, fractions are computed
both for unweighted and for completeness-weighted numbers.  We can
note that the completeness correction does not change the resulting
fractions significantly.

Except for late-types, in WINGS, fractions of different morphological
types do not depend on the choice of the mass limit, remaining
constant within the errors.
Moreover, the choice of the sample (mass or magnitude limited) does not alter  
the trends observed. 

The fraction of elliptical galaxies
does not change with redshift, being almost constant ($\sim30\%$).
Instead, the fraction of S0 and late-type galaxies considerably
changes, being S0s much more common in the Local Universe, while spirals
are proportionally less common.  

As a consequence, 
the early-type fractions strongly evolves, almost doubling from high-
to low-z. 
This is in agreement with the finding of \cite{kovac09} for the field, 
but it is in contrast with \cite{holden07}, who find
no evolution in the early-type fraction in clusters from
$z\sim0.8$ to the current epoch in a mass limited sample with $\log M_{\ast} / M_{\odot}
\geq 10.55$. Our trend still remains outstanding (
$\sim 50\%$ at z$\sim 0.8$ and $\sim 83\%$ at z$\sim 0$
) even if we adopt the \cite{holden07} mass limit.

Our results are in very good agreement with those found previously
by \cite{poggianti09} and \citet{desai07} for the WINGS and
EDisCS datasets, respectively, listed in \tab\ref{liter}.

We note that here we use only spectroscopic members in a subset of 21 WINGS
clusters, while \cite{poggianti09} used all galaxies in the photometric
catalog of 72 WINGS clusters.
Moreover, we note that considering EDisCS photo-z members,
our fraction are in perfect agreement with those found by
\cite{desai07}, who used the total photometric catalog and
a statistical background-subtraction technique,
indicating that photo-z techniques to assign membership
and statistical subtraction give consistent results.

\begin{table}
\centering
\begin{tabular}{ccc}
\hline
	& WINGS 			& EDisCS \\ 
	& $M_{V} \leq -19.5 $ 	& $ M_{V} \leq -20$ \\
	&\citep{poggianti09} 	& \citep{desai07} \\
\hline
ellipticals	& 33 $\pm$ 7\%		&29 $\pm$ 1\%	\\
S0s	& 44 $\pm$ 10\%		&14 $\pm$ 1\%	\\
early-types & 77 $\pm$ 12\%		&43 $\pm$ 2\%	\\
late-types& 23 $\pm$ 9\%		&57 $\pm$ 3\%	\\
\hline
\end{tabular}
\caption{Morphological fractions obtained from \citet{poggianti09, desai07}. \label{liter} }
\end{table}

\subsection{In what type of galaxies is most of the mass?}

In \tab\ref{massa}  we present the amount of mass in each 
morphological type, to
quantify the relative contribution of each class to the total stellar
galaxy mass.  We stress that only galaxies more massive than our
mass limits are considered, and no attempt is made to
extrapolate to lower galaxy masses to obtain the total mass,
integrated over all galaxy masses.

Errors are calculated with the propagation of errors, 
assuming a mean error on each galaxy mass equal to the mass itself,
which corresponds to the 0.3 dex error in mass discussed in \S3.

At low redshift, for $\log M_{\ast} /M_{\odot} \geq 10.2$, $\sim 85$\% of the mass is in early-type galaxies, of which
half in S0s and half in ellipticals.
At high redshift, only about 50\% of the mass is in early-type
galaxies, of which most ($\sim 3/4$) is in ellipticals.

Interestingly, the fraction of stellar mass in WINGS S0s is the same
as in EDisCS late-types.  So it is not just the number fraction of
high-z spirals (S0s) that matches that of low-z S0s (spirals), but also
the mass fractions. Consequently, it is not just the number fraction of
ellipticals that remains constant with redshift (see previous
section), but also the fraction of mass in ellipticals, being about
40\% both in EDisCS and in WINGS.

\begin{table}
\centering
\begin{tabular}{lccc}
\hline
& \multicolumn{2}{|c|}{WINGS} 		& EDisCS \\ 
& $M_{\ast} \geq 10^{9.8} M_{\odot}$ & $M_{\ast} \geq 10^{10.2} M_{\odot}$& $M_{\ast} \geq 10^{10.2} M_{\odot}$ \\
\hline
ellipticals	&42 $\pm$ 4\%	&43$\pm$ 5\%&	39 $\pm$ 5\%\\
S0s	&42 $\pm$ 4\%	&42 $\pm$ 4\%&	13 $\pm$ 2\%\\
early-types &84 $\pm$ 6\%	&85 $\pm$ 7\%&  	 52 $\pm$ 5\%\\
late-types&16 $\pm$ 2\%	&15 $\pm$ 2\%&	46 $\pm$ 4\%\\
unknown	&0.63 $\pm$ 0.22\%&0.54 $\pm$ 0.24\%&2 $\pm$ 1 \%	\\
\hline
\end{tabular}
\caption{Fraction of mass in galaxies of each type,
above the different limits for WINGS and EDisCS. 
Errors are calculated with the propagation of errors, 
assuming a mean error on each galaxy mass equal to the mass itself. \label{massa}}
\end{table}

\section{discussion}
In this paper, we have presented mass functions both of all galaxies
and of different morphological types in clusters at different epochs.

For a high mass threshold ($\log M_{\ast} / M_{\odot} \sim 11 $), 
the shapes of our $z=0.7$ and $z=0.05$
mass functions are similar, consistent with
high-mass galaxies having already established their mass
at high-z. The K-S test supports this assumption, giving a probability $\sim90\%$
that the two distributions are not driven from the same parent distribution.

This is in agreement with results
for the field, where
for example, \cite{bundy06} and \cite{franceschini06} found that
the high-mass end of the mass function is quite constant at least from
$z\sim1$ to $z\sim0$. This is also in agreement with previous studies
of the bright end of the galaxy luminosity function in clusters at red
wavelengths
(De Propris et al. 2007, Toft et al. 2004, Andreon 2006).

In contrast, going to less massive galaxies, the differences
increase: clusters in the local Universe have proportionally many more
less-massive galaxies than clusters in the distant Universe. 

We have found the same   when the sample is divided by
 morphological type. Furthermore, we have shown how 
the morphological mix changes with galaxy stellar mass,
and with redshift at a given mass.

Both the number fraction and the mass fraction of
ellipticals does not change with $z$, while the fractions of S0s and
spirals are inverted at the two redshifts. 

At the end of \S 5.1 and 5.3 we have identified
some mechanisms that in principle could contribute to
the observed evolution of the mass functions and of the
morphological mix.

In this section, we discuss each one, trying to understand
which ones can be the most relevant in driving the trends observed.

\begin{itemize}

\item {\it merging}: a galaxy can gain mass through mergers with other
galaxies; so mergers could drive both galaxy mass assembly and
dynamical evolution in a hierarchical scenario.  However, in clusters
this process is not expected to play an important role (see
e.g. \citealt{ostriker80, makino97, mihos04}): the high velocity dispersions
in clusters suppress mergers 
 (see
e.g. \citealt{ellison10} and references therein). More importantly, 
this type of process should favor the growth of
the high-mass end of the mass function relative to the low-mass end, as
we observe. Finally, visual morphological classifications show that
the fraction of colliding galaxies in clusters is low (e.g. \citealt{desai07}).

\item {\it harassment}: galaxies in clusters are subject to frequent
high speed encounters that strip a galaxy of part of its mass and
drive a morphological transformation.  Harassment has the potential to
change any internal property of a galaxy within a cluster including
the gas distribution and content, the orbital distribution of stars
and the overall shape \citep{moore96}.  Its strength depends on the
collisional frequency, on the strength of the individual collisions,
on the cluster's tidal field and on the distribution of the potential
within galaxies. It has long timescales since it needs multiple
encounters to be efficient in removing gas and stars and quenching
star formation.  It effectively perturbs low-mass galaxies while its
effects on massive objects should be less pronounced
\citep{moore96,boselli06}.

\item {\it environmental mass segregation of infalling galaxies:} clusters are far
from being closed boxes, so the infall of galaxies from surrounding
areas is a mechanism that certainly takes place. 
If galaxies surrounding the clusters have a
steeper mass function than the cluster, 
they would make a greater contribution to the total
mass function at intermediate to low-masses.

Characterizing this process is very complex, since there are several 
factors that must be understood.
 a) First
of all, we should know the rate of infalling galaxies with time, to
understand how the number of galaxies in clusters increases with
redshift. b) Then, we should know the relative fraction of the different
morphological types of infalling galaxies. It depends on the
morphological mix of galaxies located in the surrounding areas the
clusters.  In the field the morphological mix evolves with redshift
\citep{pannella06}, so we also need the proportions of the morphological types
of infalling galaxies to change from {\it z} $\sim$0.8 to {\it z} $\sim$0
(though \cite{vanderwel07} find that the early-type fraction in the
field and group environments is quite constant ($\sim$45\%) from
$z\sim0.8$ to the current epoch).  In the outskirts of the cluster,
the global morphological mixing is correlated with local density
(e.g. \citealt{huertas09}).  In general, in the field there are more
late-type galaxies than early-types (at $z\sim0.6$, $\sim 70\%$ at low
masses, $\geq 50\%$ at intermediate masses \citep{oesch09}) so late
type galaxies are probably the favourite infalling galaxies.
c) 
Most importantly, we should know the exact  
mass distribution of infalling galaxies,
that can be expected to differ from that of galaxy clusters. 
For example,
\cite{baldry06} and \cite{bolzonella09} found a dependence of the mass function
on the local density. However, as we will show later, 
comparing field and cluster galaxy mass functions, 
there seem to be no large differences 
in the different environments (see \S \ref{fi}).

\item  {\it star formation}: galaxies can increase their stellar mass
forming new stars. The fraction of star-forming galaxies is of course
much lower in clusters than in the field at the same redshift \citep{finn10}.
At least at intermediate redshifts, 
the star formation rate in star-forming cluster galaxies is similar to that of
field star-forming galaxies of the same mass, except for
a 10\% population of cluster star-forming
galaxies with reduced star formation rates
(Vulcani et al. 2010).
For the field, \cite{oesch09} have found that the
latest types of galaxies (bulge-less and irregulars) show
a non-negligible mass growth over the
time span from $z\sim0.9$ to $z\sim 0.3$, while in 
earlier types internal
star-formation since $z\sim0.9$ only results in a 
negligible increase (<1\%) in the stellar mass.

\item
Finally, when considering the evolution in the mass functions
of each morphological type, it is relevant to keep in mind that galaxies
can be subjected
to a {\it morphological transformation}, that is a change in their
morphological type, either simply due to passive evolution
or due to influences by other processes. 
Infalling galaxies meet a new environment
that can alter their properties and they are subject to several
mechanisms that e.g. can quench their star formation processes.

\end{itemize}

From these descriptions, we can conclude that merging does not play
an important role in driving the evolution of the mass functions we observe.

In principle, harassment could modify the shape of the mass
distribution, and influence the mass function of
each morphological type, altering the galaxy morphology.  
However, 
since it is more and more efficient
at lower masses, we would expect it to 
decrease the number of low mass galaxies at lower redshifts, therefore
to produce the opposite effect of what we observe (and might have to be
``overcome'' if harassment is an important process). 
Thus, even if it could contribute in shaping
and modifying the mass function, harassment does not provide 
a satisfactory explanation for our findings.

The other mechanisms listed above could play an important role in driving
the observed evolution in clusters, so in the following, we try to
quantify separately the extent of the possible contribution of each
one.

\subsection{The role of the morphological transformations}
The infall of galaxies certainly occurs at each redshift,
however it is hard to quantify its extent. 
Hence, in this section, we wish 
to test if morphological transformations from one 
type to another, alone, can
explain the evolution of the mass function of the various morphological 
types, neglecting any other possible effect, assuming
that the rate of
morphological evolution does not depend on galaxy mass and considering
clusters as closed boxes. 
In this section we only consider mass-limited samples.

\begin{figure}
\centering
\includegraphics[scale=0.45]{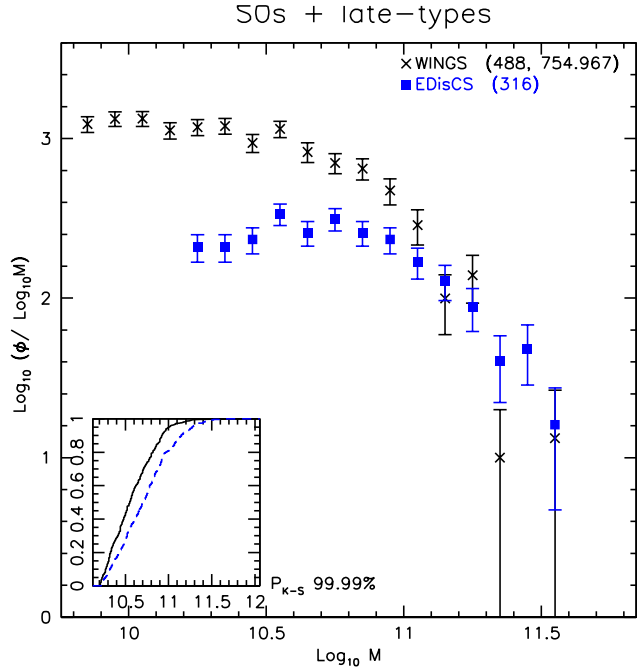}
\caption{Comparison of the mass distribution of EDisCS (blue filled
squares) and WINGS (black crosses) S0+late-type galaxies.
Mass distributions are normalized to have the same total number of
objects at $\log M_{\ast} / M_{\odot} \geq 11$. Errors are poissonian.
Numbers in the labels are the total number of galaxies above
the respective mass completeness limit. 
For WINGS, both observed (first) 
and weighted (second) numbers are given.  In the insets
the cumulative distributions are shown.  \label{S0late_norm}}
\end{figure}

Since both the mass fraction and the number
fraction in elliptical galaxies remain quite constant with
redshift (as seen also in \tab\ref{numb}), let us at first assume that 
the only
efficient mechanism is one that transforms a late-type into an S0
galaxy.  If this were the case, 
the mass distribution of late+S0 galaxies should
not change with time.  Instead, \fig\ref{S0late_norm} shows that it is
very different at the two redshifts ($P_{K-S}>99.9\%$), indicating
that the evolution of the mass functions cannot be explained simply
by turning late-type galaxies into S0s and neglecting infall and any
other mechanism.

In the next step, we recreate a more complex situation,
including elliptical galaxies in our analysis too and 
making the hypothesis that, still in a closed box, 
observed early-types in WINGS derive from
early-types in EDisCS and from a fraction of EDisCS late-types
that changed their morphology to earlier types.

In practice, since we hypothesize a closed system, 
we have to assume 
that the fraction of early-types in WINGS (84\%) 
has to be equal to the fraction
of early-types in EDisCS (47\%) 
plus some of the late-type galaxies in EDisCS necessary to maintain
the right proportions (X times the late-type fraction =
X times 58\%). We determine 
this quantity assuming that the total number 
of galaxies 
cannot change and hence X must be equal
to 0.64.

We find (\fig\ref{sim}) that the resulting simulated
mass function is very different from the observed early-type
WINGS mass function,
indicating that the early-types in WINGS cannot derive simply from
the evolution of late-types of EDisCS, no infall and no other mechanism
at work.  The K-S test support what is clear from the figure, giving 
a probability $>99.99\%$ that the distributions do not derive from
the same parent one.

Hence, it is impossible to reconcile the mass functions observed at
low-z with a simple picture in which there is no infall, each galaxy type
preserves its mass distribution unaltered
and late-type galaxies are turned into early-types by some
mechanism that leaves their stellar mass unaltered.

While morphological transformations are likely to occur,  they
have to be accompanied by other mechanisms to explain the observed trends.

\begin{figure}
\centering
\includegraphics[scale=0.45]
{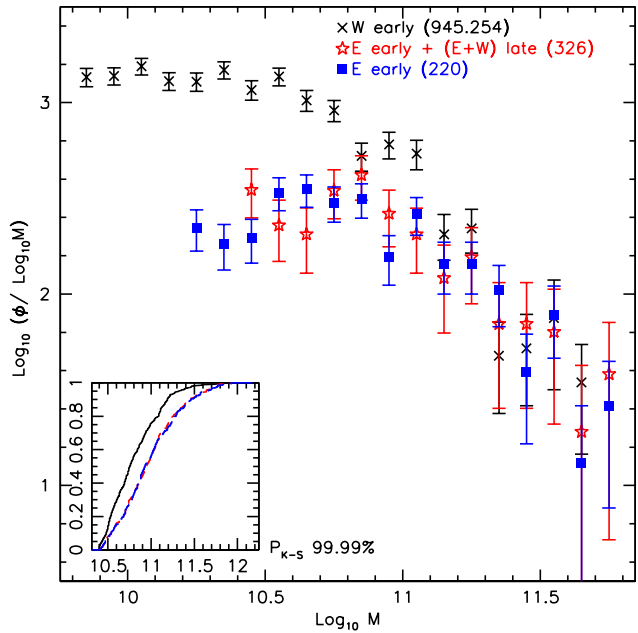}
\caption{Model of a possible evolution of the mass functions without
taking into account the infall of galaxies from areas surrounding the
clusters.  Black crosses: observed WINGS early-types, Blue filled squares:
observed EDisCS early-types (same as third panel \fig\ref{types}). Red stars:
simulated mass function of WINGS early-types
assuming they derive simply from all EDisCS early-types, and from 
some of the EDisCS late-types (see \S6.4 for details). \label{sim}}
\end{figure}

\subsection{Can infalling galaxies explain the observed evolution? } \label{fi}

Environmental  mass segregation in infalling galaxies could play an important role,
for the reasons discussed in previous sections, but we have no means to assess
its role without knowing the infall rate, mass distribution and
morphological mix of accreting galaxies.

In the literature, there are few works in which the mass
distributions of galaxies of different morphological types are
presented. \cite{ilbert10} analyzed the evolution 
from $z=2$ to $z=0$ for different morphological and spectral types in the field
using the COSMOS 2 deg$^{2}$ field; in the following,
we will consider their mass distribution of
 elliptical and all galaxies in the redshift bin $0.6\leq z\leq 0.8$. 
Bundy (2005)\footnote{From private
communication, these data are the combination of \cite{bundy05} and
\cite{bundy06}}, analyzed both together and separately, for the early- and late-type field
galaxies in the 
GOODS fields in the redshift range $0<z<1$. In the following, 
we will also consider their data
between $0.55\leq z\leq 0.8$ as reference.

First of all, we wish to compare the total mass functions.
\fig\ref{campoall} shows the comparison between cluster and field
mass functions for all galaxies, above the common mass limit, that is
$\log M_{\ast} / M_{\odot} \sim 10.3$ once all masses have been
converted to the same IMF. The mass functions are normalized in order to have
the same total number of galaxies above 
$\log M_{\ast} / M_{\odot} \geq 11$.\footnote{The number of 
galaxies per unit volume in the field
is of course much lower than in clusters, 
but here we are only interested in the shape of the distributions.}
  
We see that at high masses ($\log M_{\ast} / M_{\odot}
\geq 11$), the mass function of
field and cluster galaxies at high-z show a rather similar shape.  
However, since we are trying to understand the
reasons of the observed differences between the mass functions at
different redshifts in the lower massive tail, where WINGS and EDisCS
give different results, we are more interested in comparing the shape
of the mass function of galaxies at intermediate and low masses, assuming
the massive tail to be fixed.

As \fig\ref{campoall} illustrates, the two field studies we adopted
for comparison give quite different results at these masses. If we
followed \cite{ilbert10} (blue dashed line), we could suggest that
field galaxies have a steeper mass function than cluster galaxies,
indicating the presence of a significant environmental mass
segregation.  In this way, the infall of galaxies from the field could
be at least partly responsible for the growth of the low end of the
mass function with time.  However, even if clusters acquired an
infalling population following the \cite{ilbert10} mass function, this
would not fully explain the excess of low mass galaxies observed in
WINGS. The WINGS mass function at low masses is too high to be
produced by the field mass function observed by \cite{ilbert10} (see
the shape of the WINGS mass function, green long dashed line in
\fig\ref{campoall}).
 
Furthermore, Bundy (2005)'s results (red filled hexagons) suggest 
that there are no large
differences between the mass distribution of galaxies in the
different environments at high-z.  
In this case, an environmental mass segregation + infall surely could not be 
responsible for the observed differences between the total mass
functions in clusters at different redshifts.
Unfortunately, based on these results, we are not able to
securely establish whether 
field and cluster galaxies have similar or different mass
distributions. 

Focusing on the different morphological types (plots not shown), our
preliminary comparison between our cluster sample and the field
studies suggests that, for all types, the shape of the massive end of
the mass function does not strongly depend on environment. Instead, at
lower masses, there is a slight excess of field galaxies both for
ellipticals and for late-types; while, the comparison of early-type
galaxies (ellipticals + S0s) seems to go in the opposite direction,
with clusters having proportionally more low-mass galaxies than the
field.

We have to note that these are only preliminary results, performed
using heterogenous data and slightly different redshift ranges, so it
cannot be used to draw definite conclusions.  In the future, to
establish the existence and quantify the eventual extent of an
environmental mass segregation, it will be important to compare mass
functions of different environments at the same redshift using
homogenous data.  To understand the evolution of the mass function in
clusters, it will be especially useful to consider the mass function
in the cluster outskirts, whose galaxies will have time to become part
of clusters before $z=0$.

Nevertheless, this preliminary comparison between the cluster and field
mass functions at high-z seems to suggest that the cause for the evolution
of the mass function in clusters probably needs to be sought 
elsewhere than in environmental mass segregation effects.

\begin{figure}
\centering
\includegraphics[scale=0.45]{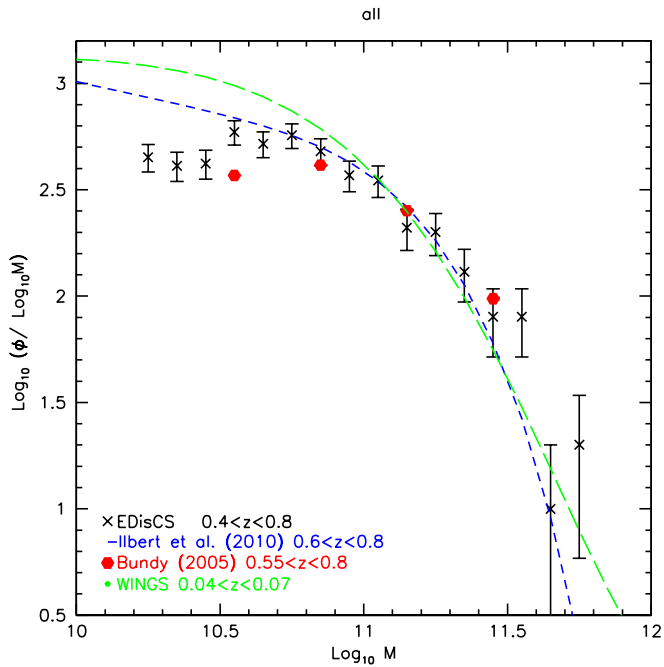}
\caption{Comparison of the mass distribution of cluster galaxies
(EDisCS - black crosses) and field galaxies (Ilbert \& Bundy) at similar 
redshifts. 
Red filled hexagons are
from \citet{bundy05} and \citet{bundy06}. We refer briefly to them as
Bundy (2005); blue dashed lines are from \citet{ilbert10}. For comparison, we also 
plot the Schechter fit of the WINGS mass function (green long dashed line).  Mass
distributions are normalized to have the same total number of objects
at $\log M_{\ast} / M_{\odot} \geq 11 $. Errors are poissonian. \label{campoall}}
\end{figure}

\subsection{How important is star formation?} \label{sfr}

We can attempt to evaluate the importance of  the
contribution of star formation in changing the mass function with
redshift, knowing the fraction of star-forming galaxies and the rate of
mass growth due to star formation.  
This process doesn't act directly by changing the
morphological type, but it increases the galaxy stellar mass.   
As previously demonstrated by several works
(see e.g. \citealt{noeske07, feulner05, perez05} and our own
\citealt{vulcani10} results), low-mass galaxies have higher values
of specific star formation rate (hereafter SSFR, star formation rate per unit
of galaxy stellar mass) than high mass galaxies so they proportionally
increase their mass much more rapidly than more massive galaxies.  As
a consequence, there can be some galaxies whose mass is below the mass
limit at the redshift of EDisCS, but with time can grow so that they
enter the sample at the redshift of WINGS, changing the shape of
the mass function at low- intermediate masses.

We build a toy model to try to quantify the importance of star formation.
First, we discuss an extreme case. 
We ``simulate'' a possible situation in which infall does not act
and the only event is the increase of mass through star formation
occurring in galaxies already located in clusters. 

In clusters, star formation cannot act undisturbed from $z\sim 0.7$ to
the current epoch, since there are probably several processes that
cause star formation to end. Therefore, we simulate a situation in which star
formation is left undisturbed only for 3 Gyr and then it stops suddenly.
  During the 3 Gyr, we hypothesize that it is not influenced
by some external processes, so it decreases naturally with time,
according to the evolution with redshift of the SSFR-Mass relation.
We have chosen 3 Gyr as a ``generous'' timescale, as this can be considered
a sort of upper limit for the time a galaxy will continue forming
stars undisturbed in clusters, 
given the dynamical timescale in clusters ($\sim 1$ Gyr)
and the long timescale expected for some of the quenching
processes proposed in clusters (e.g. strangulation, a couple
of Gyrs).\footnote{Even allowing more time for the SFR to act
(i.e. 6 Gyr), 
we found that star formation in galaxies that are already
in clusters at high-z 
cannot be the only mechanism 
responsible for the observed evolution (see below).}

\begin{figure}
\centering
\includegraphics[scale=0.45]{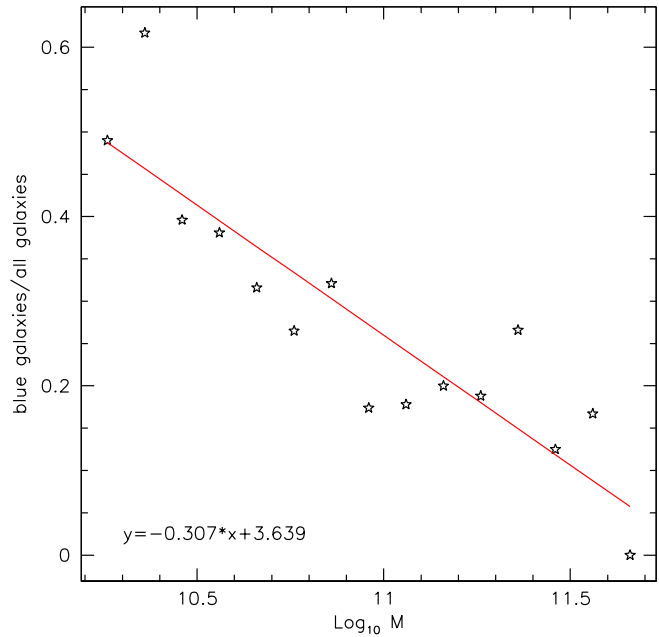}
\caption{Fraction of blue galaxies over all galaxies as a function of
the mass in the EDisCS mass limited sample.  Galaxies with $(U-V)_{rest-frame}\leq 0.8$
are considered blue, hence star-forming. The red line represents the least
squares fit performed on the points.\label{blufrac}}
\end{figure}

First of all, we take into account the fact that only star-forming galaxies
grow in mass due to star formation, 
while passive galaxies do not. We consider as
star-forming only EDisCS galaxies with $(U-V)_{rest-frame}\leq 0.8$
({\it blue galaxies}). 
For each mass, we estimate the fraction of star forming
galaxies, calculated as the number of blue galaxies over all galaxies
(\fig\ref{blufrac}).
Since the relation between the fraction of blue galaxies and $\log M$
seems to be linear, we perform a least squares fit and we 
extrapolate the fraction also at masses below the mass completeness
limit. When the fraction would turn out 
to be greater than 1, it is just set equal to 1.

In \cite{vulcani10} we analyzed the trend of the SSFR in clusters as a
function of galaxy mass. We found that this trend is very similar to
the trend found by \cite{noeske07} for the field. The main difference
is that in clusters there is a population of star forming galaxies
($\sim 10\%$) that has a reduced star formation rate for their mass
 and that probably will end the process of star formation in
the near future (for details, see \citealt{vulcani10}).  So, when we consider
the fraction of star forming galaxies we eliminate 
from the star-forming fraction  
 10\% that corresponds to galaxies with reduced star formation.

Since we need the value of median SSFR at low masses and our cluster
mass completeness limit is quite high, to estimate the mass growth we
use the median SSFR-Mass relation 
found for the field by \cite{noeske07}, calculated
in mass bin of 0.1 dex. To reach even smaller masses,
we extrapolate this relation 
down to  $\log M_{\ast} / M_{\odot} \sim 8.1$
(the minimum mass of a galaxy at $z\sim0.7 $ that it turns out
at $z\sim0$ could enter our sample due to star formation). 
We estimate by how much galaxies can
increase their mass in 3 Gyr from $z \sim 0.7$ 
adopting the SSFR appropriate for the epoch and the mass
considered. In our estimates, we account for the 
fact that the observed stellar mass at any time is equal to the
integral of the mass of stars formed before, minus the 
fraction of mass returned by stars into the interstellar phase
(about 25-35\% of the total mass every time the galaxy doubles its mass).

Finally, we assume that the EDisCS total mass function at masses below our
mass limit remains flat, i.e. has a
similar behavior to what we observe between $\log M_{\ast} / M_{\odot}=10.2$
and $\log M_{\ast} / M_{\odot}=10.8$. To verify this 
assumption where we could, 
we considered blue star-forming galaxies, for which we can reach a 
 mass completeness limit lower than that of all galaxies, that is 
 $\log M_{\ast} / M_{\odot}=10$.0.
Using the relation given in \fig\ref{blufrac}, we then estimate 
the total number of galaxies at these low masses and we find 
the result to be consistent with
our hypothesis that the EDisCS total mass function continues flat to 
low masses, at least down to $\log M_{\ast} / M_{\odot}=10$.0.

According to the fractions of star forming galaxies, we add in each
mass bin of the observed EDisCS total mass function, the number of
galaxies that could enter that bin at $z\sim 0$, and we subtract the
number of galaxies that left that bin with time. Then we normalize the
new mass function so that the number of galaxies at high masses
($\log M_{\ast} / M_{\odot}=11.5 $, where the growth turns out to be
negligible), is the same as that in the EDisCS mass function.

\begin{figure}
\centering
\includegraphics[scale=0.45]{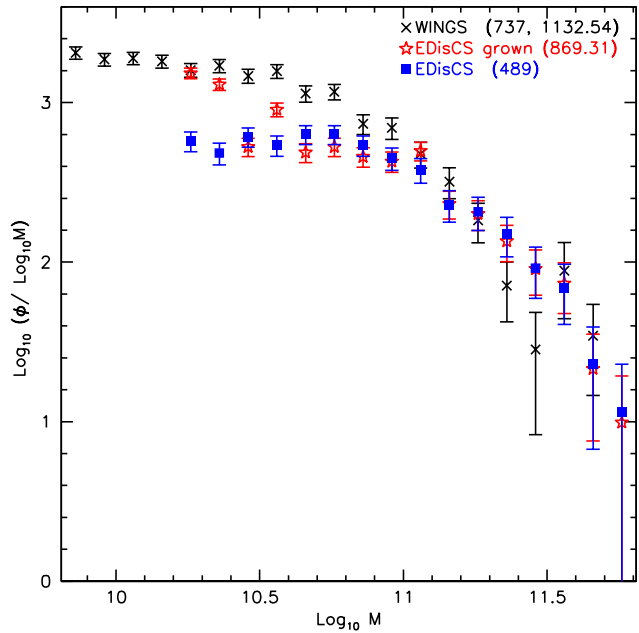}
\caption{Comparison of the mass distribution of WINGS (black crosses)
and EDisCS modified under the assumption that star formation produces
an increase at
intermediate masses (``EDisCS grown'', red empty stars).  For
reference, also the EDisCS mass function is plotted (blue filled
squares). WINGS and EDisCS mass functions are normalized as usual. The mass
distribution of EDisCS grown is normalized to have the same total
number of objects at $\log M_{\ast} / M_{\odot} \geq 11.5$ as
EDisCS. Errors are poissonian.  Numbers in the labels are the total
number of galaxies above the respective mass completeness limit.  For
WINGS, both observed (first) and weighted (second) numbers are given.
\label{provasfr}}
\end{figure}

\fig\ref{provasfr} shows the results of our attempt: we find that
high-z cluster members whose mass is increased due to star formation
could only partially fill the observed gap between WINGS and
EDisCS at low/intermediate masses. In fact, this test
could fully account for the growth of the mass function only in the
first bins, for $\log M_{\ast} / M_{\odot} \leq 10.4$, while the
growth of intermediate masses still remains unexplained.
 
We stress that this result is found assuming that the high-z mass
function is flat at masses below our limit, that infall is negligible
and that star formation continues undisturbed for only 3 Gyr in the
cluster environment. 

Then, as a second more realistic test, we also take into account the
contribution of infalling galaxies from the field, using \cite{bundy05}
data.  Since their data are given in mass bins of 0.3 dex, we use
their same binning. 
 As we do not know the infall rate, we cannot
estimate the exact relative contribution of cluster and field galaxies.
So, we normalize both the total mass function of EDisCS and that of
the field to 1 and we sum them together, giving a double weight to the
field galaxies. This correspond to the amount of cluster mass growth
between the two redshifts from dark matter simulations (Table~4 in 
Poggianti et al. 2006).
Then we assume that the star forming fraction at each mass is the same
as that used in the previous test.\footnote{We can't estimate the real
star forming fraction - mass relation in the field, hence we consider
the relation we have found for clusters. Certainly, this is a hard lower
limit and allows us to be as conservative as possible.} To extrapolate 
the field mass function at lower masses than the Bundy's
completeness limit, 
we perform a Schechter fit of the field mass function. 
To do this, we use only Bundy's late-type galaxies since
 we assume that the great majority of 
star-forming galaxies belongs to this morphological type.
 Afterwards,
in agreement with the previous toy model, we compute the mass growth
of galaxies (both of those already in clusters and of those infallen)
due to star formation in the lapse of time of 3 Gyr and estimate the
new ``grown'' mass function.

\begin{figure}
\centering
\includegraphics[scale=0.45]{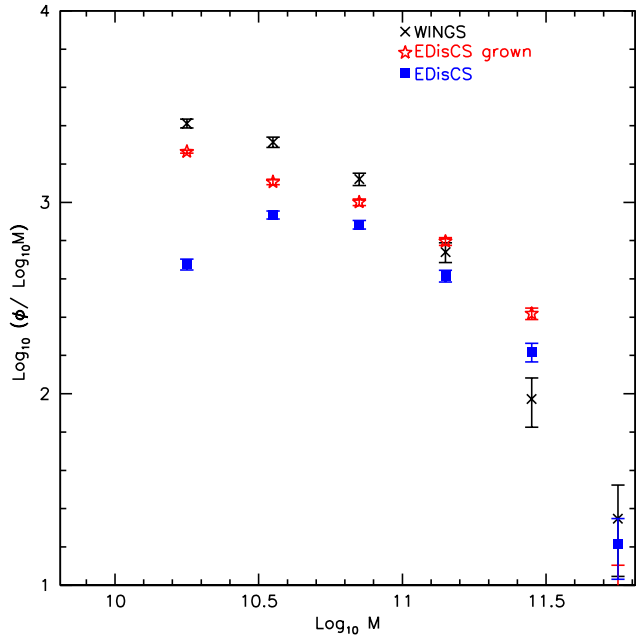}
\caption{Comparison of the mass distribution of WINGS (black crosses)
and EDisCS modified considering the contribution of the infalling
galaxies and under the assumption that SF increases the mass of
intermediate mass galaxies (``EDisCS grown'', red empty stars).  For
reference, also the EDisCS mass function is reported (blue filled
squares). 
Mass functions are normalized as usual. 
Errors are poissonian.
 \label{provasfr2}}
\end{figure}

Results are shown in \fig\ref{provasfr2}.
Comparing the shape of the ``grown'' mass function, we find that it is
more similar to the WINGS mass function than to the EDisCS one, from
which it is derived. Our result indicates that the SFR can indeed give a 
non-negligible contribution to the evolution of the mass function.
In this test, infalling galaxies are able to fill the
gap also at intermediate masses, while this did not occur
in the previous test. Importantly, the shape
of the new mass function is more similar to that of WINGS than to EDisCS.

We emphasize that in our model we have used for the field the 
star-forming fraction versus mass relation observed for clusters.
This is surely a conservative lower limit, and therefore
results in a lower limit for the contribution of infallen galaxies.
We also point out that we are supposing that all field galaxies
enter the clusters at $z\sim$0.7, while galaxies enter at
different epochs, and those entering later will have more time to continue
forming stars and contribute to the growth of the mass function. 

Moreover, we do not, for example, consider the effect of the 
enhanced star formation (starbursts) that might to be associated 
with galaxy infall into clusters \citep{dressler09}. This would provide an
additional mass growth that helps reconcile the EDisCS and WINGS
mass functions.

From these models, we suggest that star formation 
in both cluster galaxies
and, most of all, in later infalling galaxies is probably the dominant
mechanism responsible for the evolution of the total mass
function. We have seen that star formation in star-forming galaxies 
that are already in clusters at high-z does not fully explain
the observed gap in the mass functions at the two redshifts, while adding
the infalling field population more closely reproduces
the observed mass function at low-z.

This shows that infall does play a role, contributing a large number of
low mass galaxies that, as long as they continue forming stars, increase 
their mass due to star formation. In \S7.2 we have seen that it is
currently unclear whether, in addition to this effect, infall
also contributes to the evolution of the mass function via
environmental mass segregation, by directly increasing the relative 
numbers of low-mass  
($\log M_{\ast} / M_{\odot} = 10.2-10.8 $) versus high-mass galaxies
$\log M_{\ast} / M_{\odot}\geq 10.8$.

In our test we have conservatively adopted the mass function of \cite{bundy05} 
that, as we have seen in \S7.2, is  more similar to the EDisCS mass function
and suggests no environmental mass segregation. If the 
field 
mass function was steeper at lower masses 
(as \cite{ilbert10} find), the contribution 
of the mass growth due to star formation in infalling galaxies could be even 
stronger.

Obviously, star formation can affect the evolution of the late-type 
galaxy and total mass functions, but alone cannot explain the observed
evolution of the mass function of early-type galaxies. Moreover, star
formation cannot be directly responsible for the observed evolution in
the morphological mix.  Some other mechanism/s must be at work
transforming late-type galaxies into early-types, thereby changing the
shape of the early-type mass functions.\footnote{We note that such
mechanism can in principle be itself mass-dependent, and could
transform low-mass galaxies more efficiently than high-mass
galaxies. In this scenario, also harassment, by stripping mass to
galaxies and making them change their morphology, could partly
contribute to the change of the mass functions type by type.}

\section{summary and conclusions} 
In this paper, 
we have characterized the stellar
mass functions and their evolution from $z=0.8$ to $z=0$ of 
two mass limited samples of galaxies in clusters
drawn from the WINGS ($\log M_{\ast}/M_{\odot}>9.8$)
and EDisCS ($\log M_{\ast}/M_{\odot}>10.2$) datasets.
We considered both the total mass function and the mass function
of different morphological types. We then analyzed the morphological 
fractions both for mass limited and magnitude limited samples.

Our main results are:

\begin{itemize}
\item
The total stellar mass distribution of galaxies in clusters evolves
with redshift.  In a mass-limited sample, clusters at high-z
have on average more massive galaxies than at low-z. Assuming no
evolution at the massive end, the population of 
$10.2 \geq \log M_{\ast} / M_{\odot} \leq 10.8$
 galaxies must have grown significantly between $z=0.8$ and
$z=0$.

\item
At low redshift, different morphological types have significantly
different mass functions, although S0s and late-type galaxies more
massive than $\log M_{\ast} / M_{\odot} = 10.2$ may have similar mass
distributions.  At high redshift our results are not conclusive.

\item
The mass distribution of {\it each} 
morphological type evolves with redshift. In all
types there are proportionally more massive galaxies at high- than at
low-z.

\item
Both for the mass-limited and the magnitude-limited samples, the
proportion of galaxies of different morphological types as a function
of galaxy stellar mass (the morphology-mass relation) 
evolves strongly with redshift, as expected given the
evolution of the morphological fractions. 
At both redshifts, in the mass limited sample,
the incidence of each type does not 
strongly depend on mass up to $\log M_{\ast} / M_{\odot} \sim
11$. Above this value, ellipticals dominate.

\item
Both considering a mass limited and a magnitude limited sample, 
and in agreement with previous magnitude-limited works
(e.g. \citealt{dressler97, fasano00, postman05, smith05, desai07,
poggianti09}), we find that the
number fraction of elliptical galaxies is almost constant with redshift
($\sim 30\%$). In contrast, the fraction of S0s and late-types considerably
changes with time (S0s $\sim 50\%$ at the current epoch, late-type $\sim 55\%$
in the distant Universe), as also reported in those other studies. 

\item
In addition to the number fraction of ellipticals, also
the fraction of stellar mass in ellipticals remains constant
with redshift, being about
40\% both in EDisCS and in WINGS. 
At high-z, 13\% of the stellar mass is in S0s, and 46\% in later types.
At low-z, the mass fractions are inverted: 42\% in S0s, and 15\% in 
late-types.

\end{itemize}

The most likely explanation of the observed evolution of the mass
functions is the mass growth of galaxies due to star formation in both
cluster galaxies and, most of all, in galaxies infalling from the
cluster surrounding areas. In this way, galaxies that are not part of 
our distant
universe sample, after a reasonable lapse of time,
can increase their mass due to star formation so that they can enter
our local Universe sample.

Therefore, infall does play a role contributing a large number of
star-forming galaxies growing their mass. In principle, infall could
also contribute by directly increasing the population of low-mass
galaxies compared to the massive population, due to infalling galaxies
having a different mass distribution from galaxies already in clusters
at high-z. However, we do not find conclusive evidence for such
environmental mass segregation in our comparison with field studies.

In the future, it will be important to compare the galaxy stellar mass
distribution in cluster cores, cluster infalling regions, groups and
the field using homogeneous data, to conclusively establish whether
and how much the total galaxy stellar mass function depends on
environment at different redshifts.

Finally, star formation, acting mainly on late-type galaxies, can directly
affect the evolution of the total and late-type galaxy mass functions,
however it cannot directly account for the observed evolution of 
early-type galaxies. Subsequent morphological transformations from
late to early types likely cause the evolution of the mass functions
of S0s and ellipticals.

\section*{Acknowledgments}
We  thank Alessandro Bressan for providing IMF conversion factors; 
Kevin Bundy for  providing us his data and support. 
We thank John Moustakas and Patricia S{\'a}nchez-Bl{\'a}zquez 
for providing us their estimation of 
stellar masses and for helpful discussions.
We thank Vandana Desai for performing the visual classification of 
WINGS galaxies.
We thank Kai Noeske and the AEGIS collaboration
for providing us their data. 
We thank the anonymous referee, whose comments
helped us to improve the quality of this work.
BV and BMP acknowledge financial support from ASI contract I/016/07/0.

\appendix
\section{The reliability of EDisCS photo-z mass functions}
Throughout the paper, for EDisCS, we have used the photo-z membership. Here
we show that, in the mass range in common, the
mass function determined from photo-z's is in agreement within the
errors with the mass function determined using only spectroscopic members and
spectroscopic completeness weights, derived similarly to what we have done for
WINGS.  The two mass functions for the mass limited sample are shown
in (\fig\ref{edsp}).  In this comparison, we are using all photo-z and
spectroscopic members, without considering their $R_{200}$ or their
morphology.

A K-S test above $\log M_{\ast} / M_{\odot} =10.6$ 
cannot reject the hypothesis
that the two samples are drawn from the same parent distribution
(K-S probability $\sim 82.1\%$). This gives us confidence in
using the photo-z's, that allow us to reach much deeper mass limits.

\begin{figure}
\centering
\includegraphics[scale=0.45]
{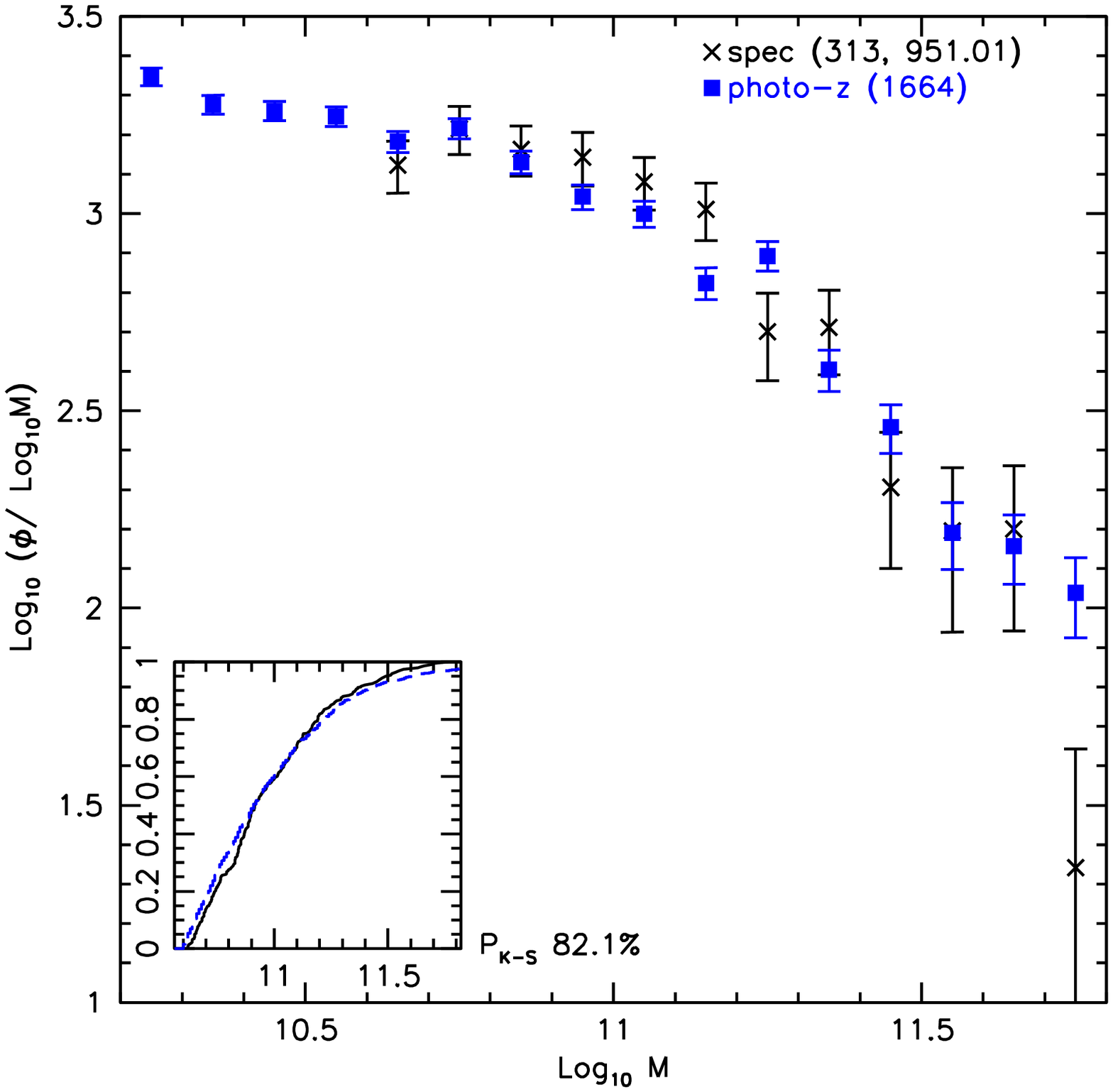}
\caption{Comparison between the mass distribution of all EDisCS galaxies
with spectroscopic membership (black crosses) and 
photo-z membership (blue filled squares). Mass distributions are normalized to have the same
total number of objects at $\log M_{\ast} / M_{\odot} \geq 11 $. Errors are
poissonian.  In the insets the cumulative distributions above the same mass limit are
shown. \label{edsp}}
\end{figure}

\section{The magnitude completeness limit}
In this Appendix we show the mass distributions in magnitude limited samples,
considering the same magnitude limits adopted by the morphological fraction
studies (\S6.1).
We show how this choice alters the final results: in fact, these 
mass distributions are very different from those we derive using
mass limited samples.

In mass selected samples only a modest and systematic evolution of
galaxy masses with time is expected, unlike magnitude selected samples
where galaxies can enter and leave the sample due to variations in
their star formation activity.  A mass selection gives a more robust
mean of identifying likely progenitors of {\it z}=0 cluster galaxies
in higher redshift clusters.  However, we note that also in mass
limited samples some progenitors could lack. First, galaxies can
continue to grow in mass and so some galaxies that weren't included at
high redshift enter in the sample at low redshift, moreover the
infall of galaxies from the field can alter the mass distribution.

It is important to note that the choice of a magnitude limit implies 
a natural mass limit below which the sample is incomplete. This corresponds
to the mass of the reddest galaxy at the limiting magnitude.

For WINGS, $M_V=-19.5$, and $B-V=1.2$, the mass completeness
limit corresponds to a $M\sim 10^{10.5} M_{\odot}$.

For EDisCS, $M_V=-20$, and  $B-V=0.9$, it corresponds to a $M\sim
10^{10.4} M_{\odot}$.

When using magnitude limited samples, it should therefore always be kept in 
mind that below the mass completeness limit the sample is incomplete, hence
results will differ from those obtained from a complete (e.g. mass limited)
sample. In the following, we show some examples. These plots have to be directly
compared with those already presented in this paper.

\begin{figure}
\centering
\includegraphics[scale=0.45]
{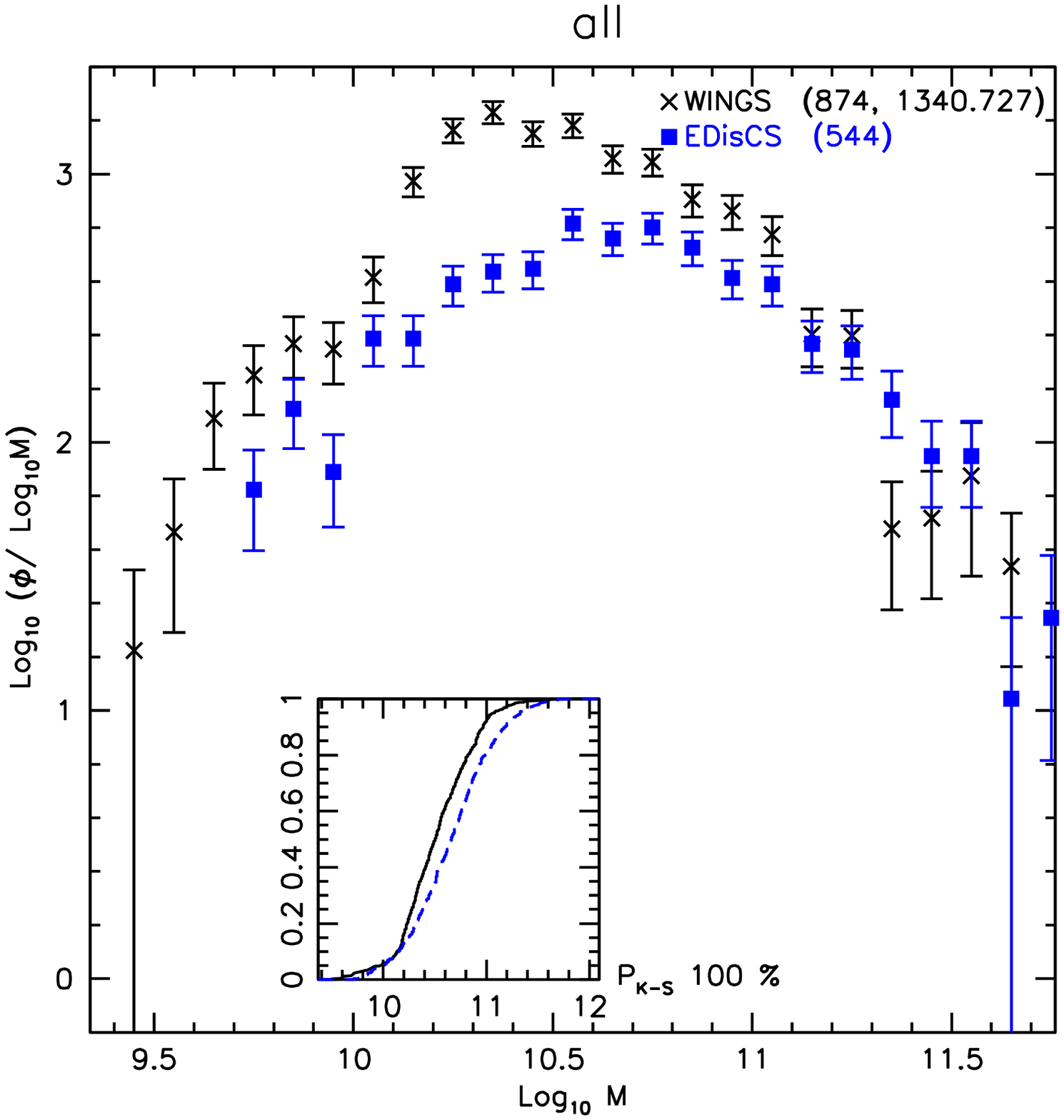}
\caption{Comparison of the mass distribution of EDisCS (blue filled
squares) and WINGS (black crosses) galaxies, of all
morphological types, for the magnitude limited samples. 
Mass distributions are normalized to have the same
total number of objects at $\log M_{\ast} / M_{\odot} \geq 11$. Errors are
poissonian.  In the insets the cumulative distributions are
shown.  \label{all_norm1}}
\end{figure}

\fig\ref{all_norm1} shows the mass distribution of all galaxies,
regardless of their morphological type, for WINGS and EDisCS galaxies
in the magnitude limited sample.  Both distributions rise, reach a peak,
then decline again.  

For WINGS, for which the completeness mass limit of the magnitude
limited sample is much higher than the limit of the mass limited
sample, it is clear that the shape of the distribution below the
completeness mass limit is very different from the one obtained for
the mass limited sample (cfr. \fig\ref{all_norm1} with
Fig.~3). Instead, no significant change is
expected nor observed between the mass limited and the magnitude
limited distributions above $ \log M_{\ast} / M_{\odot} \sim 10.5$.

As for the mass limited samples, the mass functions of WINGS and
EDisCS are very different. If we consider the whole mass range covered
by galaxies in the magnitude limited sample, a K-S test rejects the
null hypothesis of similar mass distribution with a probability
$100$\%.

\begin{figure*}
\begin{minipage}[c]{175pt}
\centering
\includegraphics[scale=0.45,clip = false, trim = 0pt 150pt 0pt 0pt]
{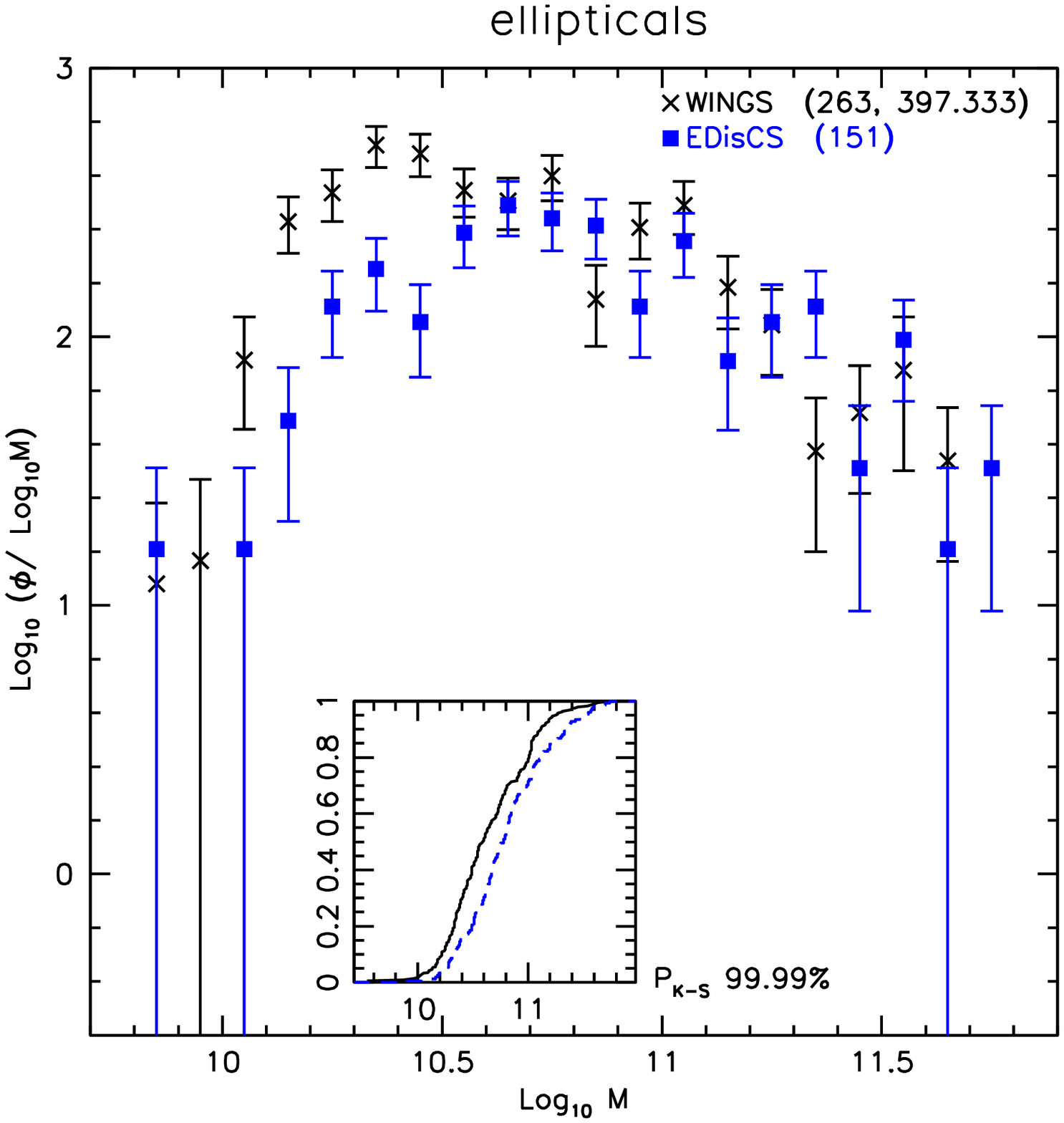}
\vspace*{2.2cm}
\end{minipage}
\hspace{2.5cm}
\begin{minipage}[c]{175pt}
\centering
\includegraphics[scale=0.45,clip = false, trim = 0pt 150pt 0pt 0pt]
{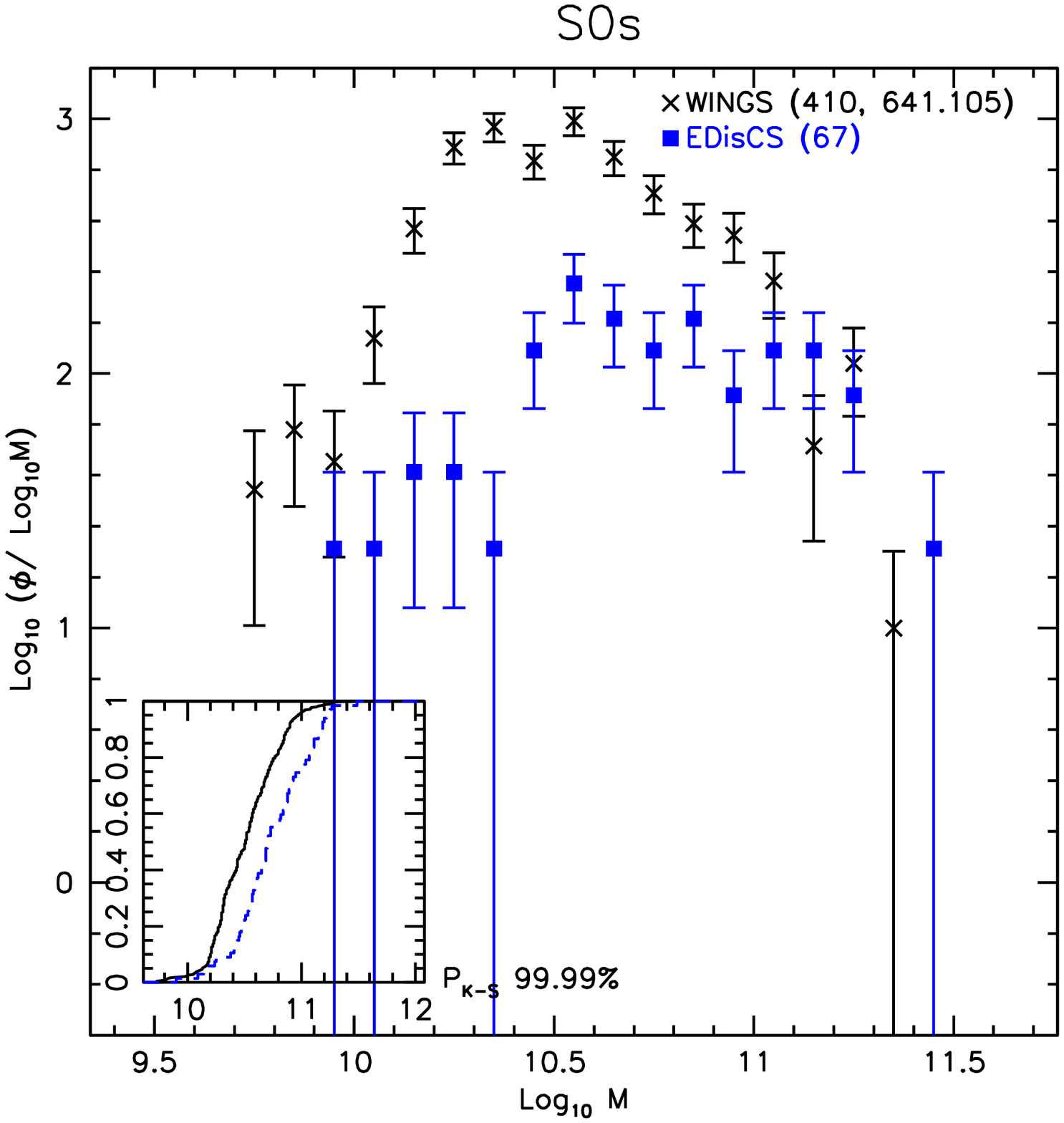}
\vspace*{2.2cm}
\end{minipage}
\begin{minipage}[c]{175pt}
\centering
\includegraphics[scale=0.45,clip = false, trim = 0pt 150pt 0pt 0pt]
{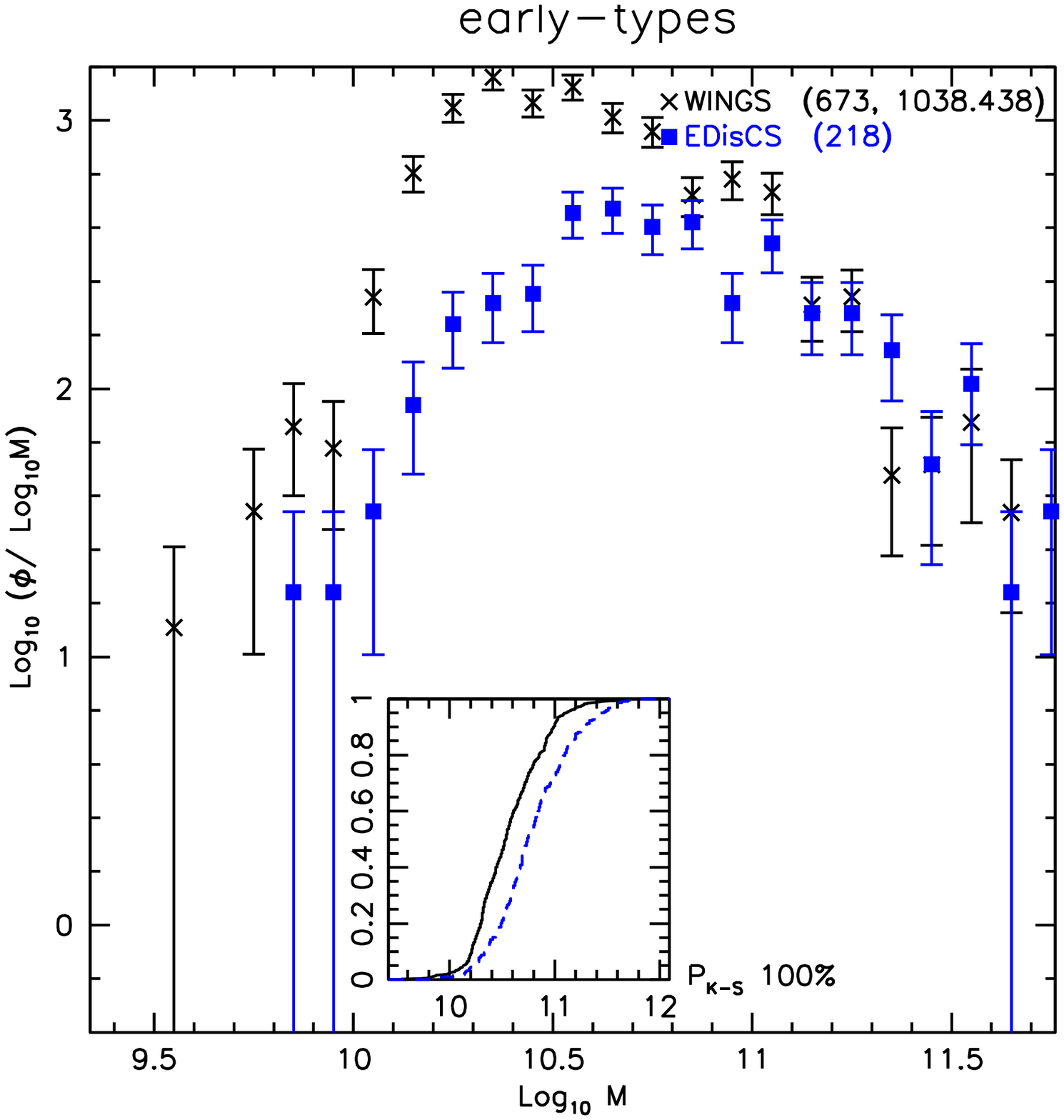}
\vspace*{2.2cm}
\end{minipage}
\hspace{2.5cm}
\begin{minipage}[c]{175pt}
\centering
\includegraphics[scale=0.45,clip = false, trim = 0pt 150pt 0pt 0pt]
{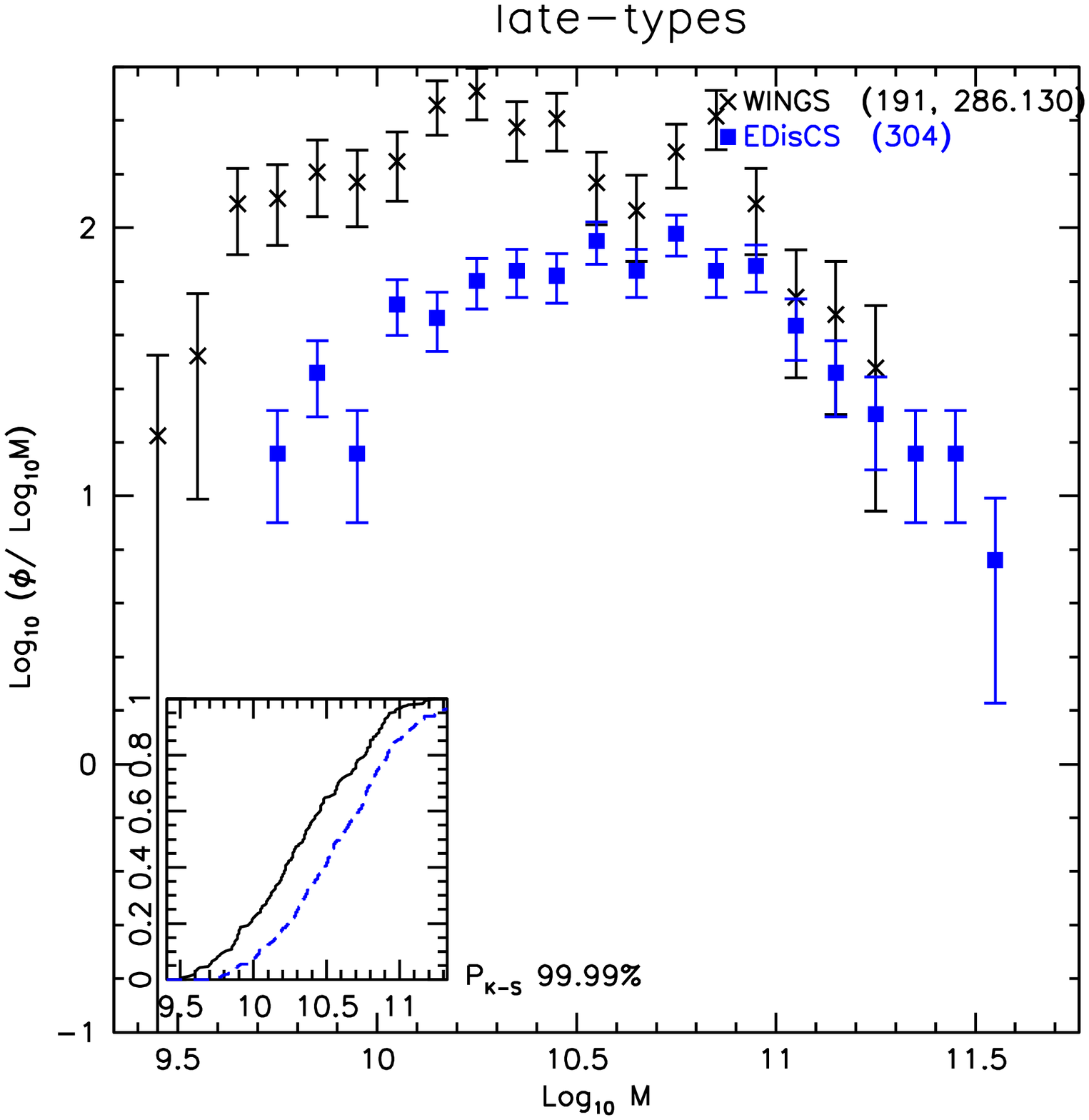}
\vspace*{2.2cm}
\end{minipage}
\hspace{1cm}
\caption{ Comparison of the mass distribution of EDisCS (blue open
circles) and WINGS (red stars) galaxies, for different morphological
types, in magnitude limited samples. Mass distributions
are normalized to the total number of objects with $\log M_{\ast} / M_{\odot} \geq 11$.
 In the insets the cumulative distributions are shown. 
\label{mag}}
\end{figure*}

Then, we also show (\fig\ref{mag}) the comparison of the mass distribution
of the different morphological types at the two redshifts, to be
contrasted with the mass-limited ones in Fig.~5.  Again, in the
magnitude limited samples, incompleteness setting in below the
completeness mass limit creates a spurious declining trend at low
masses.

The comparison of magnitude-limited and mass-limited samples clearly
shows that the mass distribution derived from the former is
meaningless, because affected by incompleteness, below the the limit
corresponding to the mass of a galaxy with the reddest color and the
faintest magnitude in the sample.

\label{lastpage}
\end{document}